\documentstyle[preprint,aps,epsfig]{revtex}
\draft
\begin{document}

\title{Transmission through a short interacting wire}

\author{D. Schmeltzer}
\address{Department of Physics, City College of the City University
of New York, New York, NY 10031}
\author{A. Saxena, A.R. Bishop, D.L. Smith}
\address{Theoretical Division, Los Alamos National Lab, Los Alamos,
NM 87545}


\maketitle

\begin{abstract}
We investigate electron transmission through a short wire with 
electron-electron interactions which is coupled to noninteracting metallic 
leads by tunneling matrix elements.  We identify 
two temperature regimes (a) $T_{Kondo}<T\le T^{wire}=\hbar v_F/k_Bd$ ($d$ 
is the length of the interacting wire) and (b) $T<T_{Kondo}\ll T^{wire}$.  
In the first regime the effective (renormalized) electron-electron 
interaction is smaller than the tunneling matrix element.  In this 
situation the single particle spectrum of the wire is characterized by 
a multilevel ``quantum dot" system with magnetic quantum number $S=0$ 
which is higher in energy than the $SU(2)$ spin doublet $S=\pm1/2$.  
In this regime the single particle energy is controlled by the 
length of the wire and the backward spin dependent interaction. 
The value of the conductance is dominated by the transmitting electrons 
which have an opposite spin polarization to the electrons in the 
short wire.  Since the electrons in the short wire have equal 
probability for spin up and spin down we find $G= G_\uparrow 
+G_\downarrow$, $e^2/h\le G< 2e^2/h$.  In the second regime, 
when $T \rightarrow0$ the effective (renormalized) electron-electron 
interaction is larger than the tunneling matrix element.  This case 
is equivalent to a Kondo problem.  We find for $T<T_{Kondo}$ the 
conductance is given by $G=2e^2/h$.  These results are in agreement 
with recent experiments where for $T_{Kondo}<T<T^{wire}$ the conductance 
$G$ obeys $e^2/h\le G< 2e^2/h$, and for $T< T_{Kondo}$, $G=2e^2/h$.  
In both regimes the current is not spin polarized and the $SU(2)$ symmetry 
is not broken.  Our model represents a good description of the 
experimental situation for an interacting wire with varying confining 
potential in the transverse direction.  
 
\end{abstract}

\vskip 0.5truecm 

\section{Introduction}
Recent experiments in quantum wires show that the spin degrees of 
freedom combined with finite size effects give rise to new interesting 
effects in the ballistic transport regime.  As an example we point 
out the new quasi-plateau with conductance $G\simeq 2(0.7$ $e^2/h)$ 
observed by the Cavendish group [1].  A number of possible explanations 
have been suggested.  Some of these explanations introduce spin 
polarization [2-4] and others are based on the Kondo effects [5,6]. 
Recently a number of groups have reported similar results.  In 
particular we mention the results obtained by Reilly et al. [7,8] 
which have reported that the conductance varies in the range 
$0.5-0.7\times2e^2/h$ as a function of the electron density, 
length of the wire and temperature.   
 
In order to clarify this problem we investigate electron transmission  
through a short wire of length $d$ with electron-electron ($e$-$e$) 
interaction coupled to non-interacting leads.  Inspired by the 
experiments in Ref. [8] we consider a short wire of length $d\sim 
0.5$-1 $\mu$m, which corresponds to a temperature $T^{wire}\simeq\hbar 
v_F/k_Bd$, where $v_F$ is the Fermi velocity and $k_B$ is the 
Boltzmann constant.  We expect that the conductance will change 
drastically when the temperature changes from $T\ll T^{wire}$ to 
$T\le T^{wire}$ \cite{david3}.  

We argue that the model of a short interacting wire coupled to 
non-interacting leads represents a good approximation to the 
experimental situation.  In the experimental case the wire widens 
smoothly from the contact region to the reservoirs.  In the contact 
region the wire is narrow, tuning the gate voltage allows a situation 
with only one propagating channel.  In the reservoirs the wire is wide 
allowing for several propagating channels. (Due to the continuity condition 
from the reservoirs region to the contact region only one channel 
in the reservoirs is a pure propagating channel and the rest have a 
complex wave vector in the wire.)  The presence of the $e$-$e$  
interaction can be separated into two parts, intrachannel and 
interchannel.  Projecting out the channels with complex wave number 
we obtain an effective one dimensional channel.  

In the reservoirs the interchannel interaction renormalizes strongly 
the propagating interacting channel.  On the other hand, in the 
wire region the interchannel renormalization is negligible.  As 
a result the reservoir region is described by a Fermi liquid contrary 
to the unrenormalized interaction in the wire region which is 
described by a one dimensional Luttinger liquid.  The transmission 
coefficient between the two regions is described by the tunneling matrix 
element $\lambda$ which couples the one dimensional Fermi liquid in the  
leads to the one dimensional Luttinger liquid in the wire.  
The matrix element $\lambda$ is further reduced by 
projecting out the non-propagating channels. 
 
To solve this transport problem we construct the effective 
spectrum of the short wire.  We find that the spectrum is built from 
charge-spin collective excitations and fermionic particle excitations. 
Using the renormalization group (RG) we find that the spectrum of the 
short wire is controlled by the zero mode fermionic single particle 
states.  The bosonic degrees of freedom are integrated out giving rise 
to an effective model for the short wire.  As a result we obtain that 
transmission across the short wire is equivalent to transmission through 
a ``multilevel state" at temperatures $T\le T^{wire}$.  The effective 
model describes the physics of length scale longer than $d$ and 
contains new exchange terms generated by the interaction of the bosonic 
degrees of freedom of the short wire.  We find that, since the 
conductance is given as the sum of the conductances for the different 
levels [see eq. (15b)], the $S=0$ state is higher in energy 
than the doublet $S=\pm1/2$, i.e. the $SU(2)$ doublet dominates the 
conductance and $G\sim e^2/h$.  

For temperature $T\rightarrow0$ so that $T<T_{Kondo}<T^{wire}$ the 
problem can be mapped into a Kondo problem which gives rise at $T 
\rightarrow0$ to perfect transmission with a conductance $G=2e^2/h$. 
This result is obtained only if the effective electron-electron 
interaction is larger than the effective tunneling matrix element. 
Therefore, this result is sensitive to temperature and length of the 
wire $d$.  

The plan of this paper is as follows.  In the next section we present 
the model relevant to a short interacting wire coupled to two 
noninteracting leads.  In Sec. III we obtain an effective model at 
(a dimensionless logarithmic) length scale $l>l_d=\log(d/a)$, where 
$d$ is the length of the interacting region and $a$ is the 
inter-electron distance. In Sec. IV we present the non-universal 
conductance at intermediate temperature and find that $G\simeq e^2/h$.  
Section V presents the computation of the conductance in the low 
temperature regime, $T<T_{Kondo}$, for which we find perfect 
transmission with the conductance $G=2e^2/h$. Finally, we summarize 
are main findings in Sec. VI.  Calculational details are relegated 
to an Appendix. 

\section{Model} 
We consider a model of two non-interacting metallic leads coupled to a 
short  interacting wire by tunneling matrix element, 
$$H=H^{leads}+H^{wire}+H_T . \eqno(1a) $$ 
The method used in this paper is as follows: (a) The two reservoirs 
are described by two one dimensional non-interacting chiral fermions. 
(b) The one dimensional interacting chiral fermion describes a short 
one dimensional interacting region restricted to $d\sim 0.5$ - 1 $\mu$m.  
(c) The transmission between the two regions is described by a matrix 
element $\lambda$ which couples the leads to the wire.  

Such a model represents the experimental situation in which  
an electronic waveguide is confined to the region $-L/2\le x\le L/2$
and $|y|\le D(x)$, where $D(\pm L/2)=W$ is the width in the
reservoir and $D(|x|\le d/2)=D_0$ is the width in the wire region
with the conditions $L\gg d$ and $W>D_0$.  In the absence of $e$-$e$
interaction this problem is solved within the Born-Oppenheimer
approximation (see Ref. [10]).  One finds that in the $y$ direction
we have a square well with transversal energies $E_n(x)=(\hbar^2/2m)
[n\pi/2D(x)]^2$, $n=1,2,...$, where $n$ corresponds to the index of
the channel.  In addition, we have a negligible matrix element $Z_{n,m}
(D(x))$ which couples the channels.  The value $D_0$ determines the
number of conducting channels $n_{max}=2k_F D_0/\pi$, where $k_F$ is 
the Fermi wave vector.  For simplicity
we consider the case with $n_{max}=1$.  This is obtained by tuning the
gate voltage $\mu^{wire}$.  Due to the fact that the width $D(x)$ 
varies in the $x$ direction, the transversal energy acts as a one-body 
potential, $E_n(x)$, and gives rise to backscattering.  In the presence 
of a 1D Luttinger liquid we expect that the scattering potential $E_n(x)$ 
will give rise to a low transmission coefficient between the two regions.
[In the presence of $e$-$e$ interaction the non-propagating interaction
channels and the matrix element $Z_{n,m}(x)$ renormalize strongly
the properties of the propagating channels in the reservoir.  We know
that for the case of coupled Luttinger chains the presence of small
matrix elements between the channels is enough to give rise to a Fermi
liquid.  Therefore we expect that such a renormalization will take
place in the reservoirs.  Consequently the reservoirs
can be described by a one dimensional Fermi liquid.  In the wire  
region this renormalization is absent and we are left with a one
dimensional interacting Luttinger liquid.  The interaction will reduce
the transmission coefficients between the two regions (Fermi
liquid-Luttinger liquid).]  Therefore we can describe the scattering
matrix between the two regions by a matrix element $\lambda<1$, see
Eqs. (6a) and (6b).

The solution of our model, with leads coupled to a short wire, is obtained 
as a linear combination of the basis chiral operators $c_{R,\sigma}$, 
$c^\dag_{R,\sigma}$ (right leads), $c_{L,\sigma}$, $c^\dag_{L,\sigma}$ 
(left leads) and $\chi_\sigma$, $\chi^\dag_{\sigma}$ (short wire). 
The fermionic operators $c_{R,\sigma}$, $c^\dag_{R,\sigma}$; 
$c_{L,\sigma}$, $c^\dag_{L,\sigma}$ and $\chi_\sigma$, 
$\chi^\dag_{\sigma}$ are constructed in each region separately for 
the case $\lambda=0$.  The chiral fermion operators are obtained as 
a product of the zero mode fermion field and an exponential of the 
particle-hole bosonic fields.  The bosonic field is periodic in each 
region.  The boundary consitions are controlled in each region by the 
chemical potentials $\mu^{\sigma}_L$ (left leads), $\mu^{\sigma}_R$ 
(right leads) and $\mu^{wire}$ (short wire). 

We construct an effective model for (a dimensionless logarithmic) length 
scale $l>\log(d/a)$, where $d$ is length of the interacting region 
$d\sim 1$ $\mu$m and $a$ is the inter-electron distance in the wire.  
Typical values of $d/a$ are in the range $10^{2}$--$10^{3}$.  In the 
leads the inter-electron distance is different than in the wire but, 
since the leads are noninteracting, the Hamiltonian is scale invariant.  
Therefore, one can use the same inter-electron distance as in the wire.  
The only difference between the two inter-electron distances is 
incorporated by rescaling the coupling constant $\lambda$.    

This construction is performed by using the Renormalization Group.  At 
this length scale the bosonic degrees of freedom of the wire have been 
completely integrated out and the fermion fields $\chi_\sigma(x)$,
$\chi^\dag_{\sigma}(x)$ [Eq. (2c)] have been replaced by the fermionic 
zero mode $V_\sigma$, $V^\dag_\sigma$ [Eq. (2d)].  The short wire 
contains renormalized interactions $\hat{g}_s$, renormalized matrix 
elements and induced exchange interaction [e.g. Eq. (A6)].  The 
renormalized ratio $\hat{\lambda}/\hat{g}_s$ and $\mu^{wire}$ [Eq. (7c)] 
control the boundary condition at the interface between the reservoirs 
and the leads.   For $\hat{\lambda}/\hat{g}_s<1$ we reproduce the Kondo 
boundary conditions.  In the opposite limit we obtain the non-universal 
conductance $G\simeq e^2/h$. 
 
\subsection{The leads}
The $H^{leads}=H^L+H^R$ represents the left and right leads restricted 
to $-L/2\le x<-d/2$ and $d/2<x\le L/2$, respectively.  We will replace 
in each lead the fermion operators with chiral right movers, $c_{L,\sigma} 
(x)$, $c^\dag_{L,\sigma}(x)$ (left lead) and $c_{R,\sigma}(x)$, 
$c^\dag_{R,\sigma}(x)$ (right lead) \cite{buttiker}.  This will be 
achieved by using open 
boundary conditions (OBC) for each lead.  As a result the wave function 
of the electron vanishes at $x=\pm d/2$ (i.e., the interface between the 
leads and the short wire).  The transmission between the leads and the 
short wire is described by a matrix element introduced at the interface 
between the two regions and is controlled by the chemical potential 
$\mu^{wire}$.  At the interface we will assume that the two wavefunctions 
in the two regions have an overlapping region of the order of the lattice 
constants.  The transmission coefficient between the leads and the wire 
can be adjusted by varying the matrix element, the overlapping region and 
the chemical potential $\mu^{wire}$.  (The concept is similar to the 
tight-binding model where localized states are used as a basis.  By varying 
the hopping matrix elements we can describe the physics of free delocalized 
electrons.) 

We start with the chiral fermion representation of the electron operators 
in the leads.  The electron field operators take the form: 
$$c_{L,\sigma}(x)=e^{ik_F(x+d/2)}L_\sigma(x+d/2)-e^{-ik_F(x+d/2)}L_\sigma 
(-(x+d/2)); ~~~~-L/2\le x<-d/2.  \eqno(1b)$$ 
Here $c_{L,\sigma}(x)$ is the fermion in the left lead expressed in 
terms of the chiral right moving fermion $L_\sigma(x)$.  Similarly, for 
the fermion in the right lead we have: 
$$c_{R,\sigma}(x)=e^{ik_F(x-d/2)}R_\sigma(x-d/2)-e^{-ik_F(x-d/2)}R_\sigma 
(-(x-d/2)); ~~~~d/2<x\le L/2,  \eqno(1c)$$ 
where $c_{R,\sigma}(x)$ is the fermion in the right lead with the chiral 
right moving fermion $R_\sigma(x)$. 

Next, we perform the continuum limit and construct the Hamiltonian for 
the leads.  We start with the Hamiltonian for the leads on the lattice 
(for each lead we have $N$ sites):     
$$H^{leads}=-t_0\sum_{\sigma=\uparrow,\downarrow}[c^\dag_{L,\sigma}(-Na-d/2) 
c_{L,\sigma}(-(N-1)a-d/2)+...+c^\dag_{L,\sigma}(-a-d/2)c_{L,\sigma}(-d/2 
-\varepsilon)+H.c]$$ 
$$-t_0\sum_{\sigma=\uparrow,\downarrow}[c^\dag_{R,\sigma} 
(Na+d/2)c_{R,\sigma}((N-1)a+d/2)+...+c^\dag_{R,\sigma}(a+d/2)c_{R,\sigma} 
(d/2+\varepsilon)+H.c] . \eqno(1d) $$ 
Here $a$ is the lattice spacing ($\sim$ inter-electron distance), 
$\varepsilon\le a$ is the overlap region between the leads and the wire, 
$t_0$ is the hopping matrix element and $Na= L/2$ is the length of the lead.  
Equation (1d) represents the tight-binding Hamiltonian for the left lead 
with fermion operators $c_{L,\sigma}$, $c^\dag_{L,\sigma}$ and the right 
lead with $c_{R,\sigma}$, $c^\dag_{R,\sigma}$.  

The OBC causes the boundary terms $c_{L,\sigma}(-d/2)$ and $c_{R,\sigma} 
(d/2)$ to vanish.  This will occur when we substitute $\varepsilon=0$ in 
Eq. (1d).  As a result we obtain $H_0^{leads}=H^{leads}(\varepsilon=0)$. 
For $\varepsilon\ne0$ we will have a transmission term from the leads to the 
short wire and in addition a non-zero boundary term.  We express 
$H^{leads}_0$ in terms of the chiral fermions. 
$$H_0^{leads}=\hbar v_F\sum_{\sigma=\uparrow,\downarrow}\int_{-L/2}^{L/2} 
dx [R^\dag_\sigma(x)(-i\partial_x)R_\sigma(x)+L^\dag_\sigma(x)(-i\partial_x) 
L_\sigma(x)]; ~~~~\hbar v_F=2t_0a\sin(k_Fa) .  \eqno(1e) $$ 

We also express Hamiltonian (1d) in terms of the {\it even} and {\it odd} 
chiral fermions 
$$\psi_{e,\sigma}(x)=\frac{1}{\sqrt{2}}[R_\sigma(x)+L_\sigma(x)] ; ~~~~ 
\psi_{o,\sigma}(x)=\frac{1}{\sqrt{2}}[R_\sigma(x)-L_\sigma(x)] . \eqno(1f) $$ 
Following the derivation in Appendix B we replace the leads Hamiltonian (1d) by 
$$H^{leads}=H_0^{leads}+H_{BC} , \eqno(1g) $$ 
$$H_0^{leads}=\hbar v_F\sum_{\sigma=\uparrow,\downarrow}\int^{L/2}_{-L/2} dx 
[\psi^\dag_{e,\sigma}(x)(-i\partial_x)\psi_{e,\sigma}(x)+\psi^\dag_{o,\sigma}(x) 
(-i\partial_x)\psi_{o,\sigma}(x)];~~~~\hbar v_F=2t_0a\sin(k_Fa) . \eqno(1h)$$  
$H_{BC}$ is the boundary term between the leads and the short interacting 
wire.  This term takes a simplified form in the Kondo regime where the 
``symmetric state" $\psi_{o,\sigma}(x)$, $\psi^\dag_{o,\sigma}(x)$ is 
screened out [the designation of the name symmetric or antisymmetric is 
given according to the original fermions $c_{R,\sigma}(x)$ and 
$c_{L,\sigma}(x)$].  $H_{BC}$ is replaced by the antisymmetric state 
$\psi_{e,\sigma}(x)$, $\psi^\dag_{e,\sigma}(x)$.  In this limit we obtain 
for $H_{BC}$, 
$$H_{BC}\simeq-2t_0(2\sin(k_Fa))^2\sum_{\sigma=\uparrow,\downarrow} 
\psi^\dag_{e,\sigma}(0) \psi_{e,\sigma}(0) . \eqno(1i) $$    

In this paper the zero-mode method \cite{david1} is used to solve the 
transport problem.  Since the fermions in the leads are non-interacting,  
we will use the fermionic representation given by Eqs. (1a)-(1i).  For 
the short interacting wire we will use the zero-mode Bosonization [see 
Eqs. (2a), (2b)].  For the sake of completeness we will also present  
the Bosonization for the leads.  Because the leads obey $L\rightarrow 
\infty$ the common belief is that the zero mode does not play any role. 
This belief is incorrect once we add particles to the ground state, in 
particular when we couple the leads to the external reservoirs [see Eq. 
(7b)].  For this situation it was shown in Ref. \cite{aleks} that in 
order to study transport with Bosonization a nonvanishing DC current 
implies the presence of gapless modes, in analogy with the Goldstone 
theorem.  The calculation in Ref. \cite{aleks} is based on anomalous 
commutators.  Similar results are obtained using the zero mode 
Bosonization \cite{david1}:    
$$R_{\sigma}(x)=\frac{1}{\sqrt{2\pi a}} e^{-i\alpha_{R,\sigma}} 
e^{i\frac{2\pi}{L}(N_{R,\sigma}-1/2)x} e^{i\sqrt{4\pi} 
\theta_{R,\sigma}(x)}  
\equiv \frac{1}{\sqrt{2\pi a}} e^{i\sqrt{4\pi}X_{R,\sigma}(x)} . 
\eqno(1k) $$    
For $L_{\sigma}(x)$ we replace $\alpha_{R,\sigma}\rightarrow 
\alpha_{L,\sigma}$, $N_{R,\sigma}\rightarrow N_{L,\sigma}$, 
$\theta_{R,\sigma}(x)\rightarrow \theta_{L,\sigma}(x)$ and 
$X_{R,\sigma}(x)\rightarrow X_{L,\sigma}(x)$,     
where $\alpha_{R,\sigma}$, $\alpha_{L,\sigma}$ are the zero mode 
coordinates; $N_{R,\sigma}$, $N_{L,\sigma}$ are the number operators; 
and $\theta_{R,\sigma}(x)$, $\theta_{L,\sigma(x)}$ are the 
particle-hole excitations.  We have the following commutation rules, 
$[\alpha_{R,\sigma}, N_{R,\sigma'}]=i\delta_{\sigma,\sigma'}$, 
$[\alpha_{L,\sigma}, N_{L,\sigma'}]=i\delta_{\sigma,\sigma'}$.  
The zero mode contribution to the Bosonized Hamiltonian is negligible 
since in the limit $L\rightarrow \infty$, $N_{R,\sigma}/L\rightarrow0$ 
and $N_{L,\sigma}/L\rightarrow0$. 

\subsection{The short interacting wire}
Next, we consider the short interacting wire of length $d\ll L$. We will 
use the same Fermi momentum for the short wire as for the leads. 
According to the experimental situation we expect that the Fermi 
momentum in the wire region is smaller than the one in the leads 
(due to the change of the width).  This shift of the Fermi momentum in 
the wire is considered by adding to the wire Hamiltonian the chemical 
potential $\mu^{wire}$ [see eq. (7c)] which shifts the wire ground state.   

The finite size of the system and the boundary conditions for 
the short wire allow us to introduce zero mode excitations. 
The fermion $d_\sigma(x)$, $d^\dag_\sigma(x)$ on the short wire 
is represented by anti-periodic right chiral fermions $\chi_\sigma(x)$, 
$$d_\sigma(x)=e^{ik_Fx}\chi_\sigma(x)+e^{-ik_Fx}\chi_\sigma(-x) . \eqno(2a)$$ 
Here $d_\sigma(x=d/2)=d_\sigma(x=-d/2)=0$ is obtained by demanding 
anti-periodic boundary conditions for the chiral fermion in the short wire. 
$$\chi_\sigma(x+d)=\chi_\sigma(x)e^{i\pi} .  \eqno(2b)  $$
  
Using the zero mode Bosonization introduced in Ref. \cite{david1}, we obtain 
the representation 
$$\chi_\sigma(x)=V_\sigma e^{(i2\pi/d)(p_\sigma-1/2)x}e^{i\sqrt{4\pi} 
\phi_\sigma(x)}, \eqno(2c) $$ 
$$V_\sigma=\frac{1}{\sqrt{2\pi a}}e^{-iq_\sigma} \equiv \frac{1} 
{\sqrt{2\pi a}}\hat{V}_\sigma, 
~~~\sigma=\uparrow,\downarrow .  \eqno(2d) $$ 
Here $q_\sigma$ and $p_\sigma$ are the zero mode coordinate and ``momentum", 
$[q_\sigma,p_{\sigma'}]=i\delta_{\sigma\sigma'}$, respectively. The ``momentum" 
$p_\sigma=0,1,2, ...$ measures the change of fermion number with respect 
to the filled Fermi sea. $\phi_\sigma(x)$ is the non-zero mode (the particle-hole) 
bosonic excitation. The operators $\hat{V}^\dag_\sigma$, $\hat{V}_\sigma$ 
act as creation and annihilation fermion operators, $\hat{V}^\dag_\sigma 
|p_\sigma\rangle=|p_\sigma +1\rangle$.  The operators $\hat{V}^\dag_\sigma$, 
$\hat{V}_\sigma$ obey 
anticommutation relations, $\{\hat{V}^\dag_\sigma,\hat{V}_{\sigma'}\}=2 
\delta_{\sigma\sigma'}$, 
$\{\hat{V}^\dag_\sigma,\hat{V}^\dag_{\sigma'}\}= 
\{\hat{V}_\sigma,\hat{V}_{\sigma'}\}=0$ and commutation relations with the 
number operator $p_\sigma$; $[p_\sigma,\hat{V}_{\sigma'}^\dag]=\delta_{\sigma 
\sigma'} \hat{V}_{\sigma'}$; $[p_\sigma,\hat{V}_{\sigma'}]=-\delta_{\sigma 
\sigma'}\hat{V}_{\sigma'}$.  

The form and condition in Eqs. (2a)-(2c) are chosen for the following reasons: 
(a) We intend to construct the bosonic representation of the chiral fermion 
field $\chi_\sigma(x)$ which is given as a product of the zero mode fermion 
$V_\sigma \exp{(i2\pi/d)(p_\sigma-\delta)x}$ and the exponential of a 
periodic bosonic field $\phi_\sigma(x)$.  The boundary condition of the 
fermion field $d_\sigma(x)$ in Eq. (2b) is determined by the value of 
$\delta$.  In Eqs. (2b) and (2c) we have $\delta=1/2$.  The representation 
in Eq. (2c) is given for $\delta=1/2$ and has no fermions in the ground 
state, $\langle p_\sigma\rangle=0$, $p_\sigma=:p_\sigma:$, where $:p_\sigma:$ 
stands for normal order and $\langle p_\sigma\rangle$ means the expectation 
value with respect to the filled Fermi sea. 

Equation (2c) is obtained in the absence of a chemical potential of the 
wire.  Once a gate voltage $2V_G$ is applied, we must add to the 
Hamiltonian a term $\mu^{wire}(p_\uparrow+p_\downarrow)$ [see Eq. (7c)]. 
The presence of the nonzero chemical potential $\mu^{wire}$ shifts the ground 
state occupation.  Consequently $\langle p_\sigma\rangle\ne0$ and $p_\sigma 
-1/2$ is replaced by $:p_\sigma:+\langle p_\sigma\rangle-1/2$. 
The shift $\langle p_\sigma\rangle-1/2$ can give rise to a change in the 
boundary conditions.  We observe that the coupling Hamiltonian in Eqs. (9d) 
and (11c) below is controlled by $p_\sigma=\langle p_\sigma\rangle+:p_\sigma:$, 
with $\langle p_\sigma\rangle$ determined by the renormalized chemical 
potential $E_\sigma(d)$ [see Eq. (9h)].  In the intermediate range of 
temperatures (see Sec. IV) the effect of the chemical potential in the wire 
is considered by taking the expectation value with respect to $p_\sigma$ 
[see Eq. (15b)].    

The physical reason for the representation in Eqs. (2a-2c) has to do 
with the fact that at the (dimensionless) length scale $l>\log(d/a)$ the 
short wire is replaced by an impurity atom which in the limit 
$T\rightarrow0$ corresponds to a single impurity.  In this regime 
$\lambda\ll1$ and the $e$-$e$ interaction $\hat{g}_s$ is large (see Sec. 
V).  This gives rise to the Kondo physics.  

Using the chiral representation given by Eq. (1e), we obtain the Hamiltonian 
for the short wire Hubbard model 
$$H^{wire}=H_0^{wire}+H_u^{wire}+H_B^{wire},  \eqno(3a) $$ 
where 
$$H_0^{wire}=\int_{-d/2}^{d/2}dx\{\sum_{\sigma=\uparrow,\downarrow}\hbar 
\tilde{v}_F\chi^\dag_\sigma(x)(-i\partial_x)\chi_\sigma(x)+\tilde{U}: 
\chi^\dag_{\uparrow}(x)\chi_{\uparrow}(x):\chi^\dag_\downarrow(x)\chi_\downarrow(x)$$  
$$+\tilde{U}:\chi^\dag_{\uparrow}(x)\chi_{\uparrow}(x):\chi^\dag_\downarrow(-x) 
\chi_\downarrow(-x)\} \eqno(3b) $$ 
and $U$ is the original Hubbard interaction, $\tilde{U}=2U$ and $\tilde{v}_F 
=2v_F$. The Umklapp and the backward term Hamiltonians are respectively given by 
\cite{neko}: 
$$H_u^{wire}=\frac{1}{2}\tilde{U}\int_{-d/2}^{d/2}dx\{\chi^\dag_{\uparrow}(x) 
\chi^\dag_\downarrow(x)\chi_\downarrow(-x)\chi_{\uparrow}(-x)e^{i4k_Fx}+H.c.\},  
\eqno(3c) $$  
$$H_B^{wire}=\frac{1}{2}\tilde{U}\int_{-d/2}^{d/2}dx\{\chi^\dag_{\uparrow}(-x) 
\chi^\dag_\downarrow(x)\chi_\downarrow(-x)\chi_{\uparrow}(x)+H.c\}. \eqno(3d) $$    
  
Next we bosonize the Hamiltonian given in Eqs. (3a)-(3d). $H_0^{wire}$ can 
be replaced by $H_0^{(n=0)}$ (the zero mode) and $H_0^{(n\ne0)}$ (non-zero 
mode part):  
$$H_0^{wire}=H_0^{(n\ne0)}+H_0^{(n=0)}, \eqno(4a) $$ 
$$H_0^{(n\ne0)}=H_{0,c}^{(n\ne0)}+H_{0,s}^{(n\ne0)} .  \eqno(4b) $$ 
Here $H_{0,c}^{(n\ne0)}$ represents the charge density and $H_{0,s}^{(n\ne0)}$ 
is the spin density. 
$$H_{0,c}^{(n\ne0)}=\int_{-d/2}^{d/2}dx\tilde{V}_c(\partial_x\tilde{\phi}_c(x))^2 , 
\eqno(4c) $$ 
$$H_{0,s}^{(n\ne0)}=\int_{-d/2}^{d/2}dx\tilde{V}_s(\partial_x\tilde{\phi}_s(x))^2 ,  
\eqno(4d) $$ 

The renormalized bosonic fields, $\tilde{\phi}_c(x)$ and $\tilde{\phi}_s(x)$ 
(charge, spin) are related to $\phi_c(x)$ and $\phi_s(x)$ by: 
$$\phi_{c(s)}(x)=\frac{K^{-1/2}_{c(s)}}{2}\left(\tilde{\phi}_{c(s)}(x) 
+\tilde{\phi}_{c(s)}(-x)\right)+\frac{K^{1/2}_{c(s)}}{2}\left(\tilde 
{\phi}_{c(s)}(x) -\tilde{\phi}_{c(s)}(-x)\right) , \eqno(4e) $$ 
with $K_c$ and $K_s$ given by: 
$$K_c=\sqrt{\frac{1-U/\pi v_+}{1+U/\pi v_+}}, ~~~~v_+=\tilde{v}_F\left(1 
+\frac{U} {\pi\tilde{v}_F}\right) ,$$ 
$$K_s=\sqrt{\frac{1+U/\pi v_-}{1-U/\pi v_-}}, ~~~~v_-=\tilde{v}_F\left(1 
-\frac{U} {\pi\tilde{v}_F}\right) , \eqno(4f) $$ 
and charge and spin velocities 
$$v_c=\frac{\tilde{v}_F}{K_c}, ~~~ v_s=\frac{\tilde{v}_F}{K_s} . \eqno(4g) $$   

The zero mode part of $H_0$ allows viewing the short interacting wire as 
a multilevel ``quantum dot" with single particle energy given by: 
$$H_0^{(n=0)}=\frac{h\tilde{v}_F}{2d}(p^2_\uparrow+p^2_\downarrow) 
+\frac{U(4\pi)^2}{d}p_\uparrow p_\downarrow , \eqno(5a) $$ 
where $p_\sigma=0,1,2,...$ represents the additional charges with 
respect to the filled Fermi sea $|p_\uparrow=0,p_\downarrow=0\rangle$. 
Typical values considered in Ref. [8] are $d_0\sim 0.5$ $\mu$m. 
This corresponds to energies $\epsilon_0=h\tilde{v}_F/2d=10^{-19} 
(v_F/c)(d_0/d)$ Joule, since $v_F/c\sim10^{-3}$, $\epsilon_0$ is of the 
order of meV or temperature $T^{wire}=\epsilon_0/k_B=7.6\times10^3(v_F/c) 
(d_0/d)$ which corresponds to a few Kelvins.  

The Bosonization of the Umklapp and backward terms involves both the 
bosonic degrees $\phi_\sigma(x)$ and the zero modes $p_\sigma$, and 
fermionic operator $\hat{V}_\sigma$. 
$$H_u^{wire}=\int_{-d/2}^{d/2}dx\left\{\frac{U}{2\pi^2a^2}\hat{V}^\dag_\uparrow 
\hat{V}_\uparrow\hat{V}^\dag_\downarrow\hat{V}_\downarrow\cos\left[(4k_F-G)x 
+\frac{2\pi}{d}(p_\uparrow+p_\downarrow)+\sqrt{8\pi}(\phi_c(x)-\phi_c(-x)) 
\right]\right\}. 
\eqno(5b) $$ 
We find that the Umklapp term is controlled by the charged boson 
$\phi_c(x)$, the charge of the ``dot" $p_c=p_\uparrow+p_\downarrow$ and the 
reciprocal lattice vector $G=2\pi/a$.  This term is highly sensitive to the 
electron density, namely $k_F=(\pi/2a)(1-\delta)$. For density $\delta$ 
and wires of length $d$ which satisfy $(4k_F-G)d\ge 2\pi$ or $\delta(d/a) 
\gg 1$, the Umklapp term can be neglected. Since $d\sim0.5$ - 1 $\mu$m the 
deviation from half filling must be of the order less than 1\%.  Under this 
condition the Umklapp term will not give rise to a charge gap. 

Next, we consider the backward term, $H_B^{wire}$.  In the limit 
$d\rightarrow\infty$ this term renormalizes to zero driving $K_s\rightarrow1$.  
For finite $d$ this is not the case:  
$$H_B^{wire}=\int_{-d/2}^{d/2}dx\left\{\frac{U}{2\pi^2a^2}\hat{V}^\dag_\uparrow 
\hat{V}_\uparrow\hat{V}^\dag_\downarrow\hat{V}_\downarrow\cos\left[\frac{2\pi}{d} 
(p_\uparrow-p_\downarrow)x+\sqrt{8\pi}(\phi_s(x)-\phi_s(-x))\right]\right\}. 
\eqno(5c)$$ 
The background term is controlled by the bosonic spin density $\phi_s(x)$ and 
spin excitations $p_s=p_\uparrow-p_\downarrow$. 
  
\subsection{The coupling Hamiltonian} 
The coupling Hamiltonian is given by 
$$H_T=\sum_{\sigma=\uparrow,\downarrow}[-t_Lc^\dag_{L,\sigma}(-d/2-\varepsilon) 
d_\sigma(-d/2+\varepsilon)-t_Rc^\dag_{R,\sigma}(d/2+\varepsilon)d_\sigma(d/2 
-\varepsilon) +H.c] . \eqno(6a) $$ 
In Eq. (6a) we have introduced an ``overlap" $\varepsilon\sim a$ between the 
electrons in the wire and leads in order to allow transmission. (Due to the 
boundary condition when $\varepsilon=0$ we have $H_T\equiv0$).  Next we 
assume $t_L=t_R=t$ and make use of the chiral representation [see Eqs. 
(1b,1c,1f,2a)].  We find: 
$$H_T=i\lambda\sum_{\sigma=\uparrow,\downarrow}\sum_{R=\pm(d/2-\varepsilon)} 
\int_{-d/2}^{d/2}dx\delta(x)[\psi^\dag_{0,\sigma}(x)e^{ik_FR}\chi_\sigma(R) 
-e^{-ik_FR}\chi^\dag_\sigma(R)\psi_{0,\sigma}(x)] .  \eqno(6b) $$ 

The tunneling matrix element $\lambda=\lambda(\Lambda)^0$, where $\Lambda=1/a$ 
is the cutoff and $\lambda=2\sqrt{2}t\sin(k_Fa)\equiv(t/t_0)E_F\sin(k_Fa)$, 
$E_F\equiv\hbar v_F/2a$. 

The matrix element $\lambda$ obeys $\lambda<1$ (this being a
result of the $e$-$e$ interactions in the leads which have been integrated
out).  This value of $\lambda$ is fixed by the transmission strength $t/t_0$
and the overlapping region of the wavefunction $\varepsilon\sim a$.  We
considered the short wire as a localized state which is coupled to the 
leads.  Controlling the strength of $\lambda$ we can describe {\it extended}
solutions in spite of the fact that we have started from a localized
picture (this is the philosophy of the tight-binding method).

We argue that for our problem this is a good starting point since
renormalization effects further decrease $\lambda$ and give rise, for
the (dimensionless) length scale $l>\log(d/a)$, to an impurity problem 
which can be mapped into a Kondo problem for $T\rightarrow0$.

\subsection{Computation of the current} 
To compute the transmission current we add the reservoir Hamiltonian $Y$:  
$$Y=Y^{leads}+Y^{wire} , \eqno(7a) $$      
$$Y^{leads}=\sum_{\sigma=\uparrow,\downarrow}(\mu_L^{(\sigma)}N_{L,\sigma} 
+\mu_R^{(\sigma)}N_{R,\sigma}) , \eqno(7b) $$ 
with $(1/2)(\mu^\uparrow_L+\mu^\downarrow_L)-(1/2)(\mu^\uparrow_R 
+\mu^\downarrow_R)=eV_{DS}$ the voltage difference, and $N_{L,\sigma} 
=\int_{-L/2}^{L/2}dx L^\dag_\sigma(x)L_\sigma(x)$, $N_{R,\sigma} 
=\int_{-L/2}^{L/2} dx R^\dag_\sigma(x)R_\sigma(x)$ are the fermion 
densities in the leads. 
$$Y^{wire}=\mu^{wire}(p_\uparrow+p_\downarrow) , \eqno(7c) $$ 
with $\mu^{wire}=eV_G$ being the gate voltage applied to the wire.  The term 
$Y^{wire}$ in Eq. (7c) allows for a nonzero number of fermions with 
respect to the leads Fermi sea.  When $\mu^{wire}\ne0$, $k_F$ in Eqs. 
(2a)-(2c) is shifted to $k_F^{wire}\equiv k_F+(2\pi/d)\langle p_\sigma 
\rangle$ and $p_\sigma$ is replaced by $:p_\sigma:$. This procedure 
accommodates the experimental situation, $k_F^{wire}<k_F$ ($\mu^{wire}$ 
can be tuned to $\langle p_\sigma\rangle <0$ needed in the experiments). 
We note that the conductance formula given below in Eq. (15) is sensitive 
to this tuning. The current operator $I_\sigma$ is given by 
$$I_\sigma(t)=\frac{e}{2}\frac{d}{dt}(N_{R,\sigma}-N_{L,\sigma}) 
=\frac{e}{i2\hbar}[N_{R,\sigma}-N_{L,\sigma},H_T]$$  
$$=\frac{e\lambda}{2\hbar} \sum_{R=\pm(d/2-\varepsilon)}[\psi^\dag_{e,\sigma}(0) 
e^{ik_FR}\chi_\sigma(R)+e^{-ik_FR}\chi^\dag_\sigma(R)\psi_{e,\sigma}(0)] . \eqno(7d) $$ 

The non-equilibrium value of the current is obtained by performing the 
thermal expectation with respect to the grand canonical Hamiltonian, 
$\hat{H}=H-Y$, where $H$ is given in Eq. (1a) and $Y$ in Eq. (7a).  As 
a result we obtain 
$$\langle\langle I_\sigma \rangle\rangle=\frac{Tr(e^{-\beta\hat{H}}I_\sigma)} 
{Tr(e^{-\beta\hat{H}})} . \eqno(7e) $$ 
Equation (7e) will be used below for the computation of the current.  

Equation (7e) with the reservoir Hamiltonian $Y^{leads}=eV_{DS}N_L/2  
-eV_{DS}N_R/2$ [see Eq. (7b)], with $\mu_L=eV_{DS}/2$, $\mu_R=-eV_{DS}/2$ 
and $N_L\equiv N_{L\uparrow}+N_{L\downarrow}$, $N_R\equiv N_{R\uparrow}+ 
N_{R\downarrow}$ allows to compute the current taking into account the 
boundary imposed by the reservoirs.  Consequently, the density in the 
right and left leads obey the boundary conditions given in Eq. (14c).  
This method has been used in Ref. \cite{aleks}.    
In particular, it has been shown in Ref. \cite{aleks} that a one-dimensional 
interacting fermion (with no backward and Umklapp interaction) coupled 
to non-interacting leads has a universal conductance $G=2e^2/h$.  This 
result has been shown to follow from the dynamical requirements, $[N_R,N_L]= 
[H,N_R]=[H,N_L]=0$.  ($H$ is the interacting Hamiltonian constructed in 
Refs. \cite{ponom,maslov,oreg}).  The non-renormalization of the conductance 
by electron-electron interactions has been shown originally 
\cite{ponom,maslov,oreg} to be a consequence of the strong influence of 
the boundary conditions imposed by reservoirs.  The same results have been 
obtained in Refs. \cite{david1,aleks} using the $Y^{leads}$ Hamiltonian 
[see Eq. (7b)].  The advantage of the formalism used in Ref. \cite{aleks} 
and the zero mode calculation given in Ref. \cite{david1} is the simplicity.  
On the other hand the result derived in Ref. \cite{ponom,maslov,oreg} are 
based on the exact integration of the electrostatic potential $\mathcal{V}$(x) 
along the one dimensional channel $\int_{-L/2}^{L/2}\rho(x)\mathcal{V}$(x)dx 
where $\rho(x)$ is the electronic density.  For problems with backscattering 
[induced by the one body potential $E_n(x)$] and backward interaction, 
the method given in Ref. \cite{aleks} is easy to use.  The simplicity of 
the reservoir Hamiltonian $Y^{leads}=\mu_R N_R+\mu_L N_L$ (see Ref. 
\cite{aleks}) causes only the zero mode coordinates (or the Goldstone 
modes) to be affected by the external boundary conditions.  At finite 
temperature we use Eq. (7e) and at $T=0$, $Y^{leads}$ is part of the 
Hamiltonian which will affect the zero mode variables.  When $Y^{leads} 
\ne0$ the ground state is shifted, consequently the zero mode coordinates 
become $\alpha_{R,\sigma}(t) \rightarrow \alpha_{R,\sigma}(t)-eV_{DS} t/2 
\hbar$ and $\alpha_{L,\sigma}(t)\rightarrow\alpha_{L,\sigma}(t)+ eV_{DS} 
t/2\hbar$.  As a result the chiral fermion in the leads will be modified. 
The current will be computed by replacing the tunneling Hamiltonian given 
in Eq. (6b) by a time-dependent Hamiltonian.  Therefore, $R_\sigma(x)$ and 
$L_\sigma(x)$ in Eq. (6b) and (7d) will be modified: $R_\sigma(x,t) 
\rightarrow\exp(i eV_{DS}t/2\hbar)R_\sigma(x,t)$ and $L_\sigma(x,t) 
\rightarrow \exp(-i eV_{DS}t/2\hbar)L_\sigma(x,t)$. 

The boundary condition affects only the zero modes $\alpha_{R,\sigma}$, 
$\alpha_{L,\sigma}$ (leads) and $q_\sigma$ (wire).  Therefore, when we 
integrate short distance behavior given by $\phi_\sigma(x)$ in Eq. (2c),  
this has no effect on the zero modes and thus on the boundary condition.  
According to Refs. \cite{ponom,maslov,oreg} this does not appear to 
be correct since the result depends on the integral of the electrostatic  
potential $\mathcal{V}$(x) along the one-dimensional channel.  The 
resolution of this problem can be achieved by replacing the electrostatic  
potential $\mathcal{V}$(x) with an averaged electrostatic potential. Using 
this approximation will allow us to neglect the effect of the boundary on 
the short distance renormalization processes.  For the rest of the paper 
we will work at finite temperature.  We will use $Y^{leads}=\mu_R N_R+ 
\mu_L N_L$ according to Eq. (7b) and will compute the current using Eq. 
(7e) in agreement with Ref. \cite{aleks}.  By doing so we ignore the 
short length fluctuations of the electrostatic potential considered in 
Refs. \cite{ponom,maslov,oreg}.

\section{Effective model} 
The effective model at the (dimensionless) length scale, $l\ge l_d\equiv 
\log(d/a)$, $T\le\hbar v_F/k_Bd$ can be mapped to an impurity model. 
To obtain the effective model we integrate out the short 
wire degrees of freedom.  The wire field operators are given by 
$$\chi_\sigma(x)=V_\sigma e^{i(2\pi/d)(p_\sigma-1/2)x} e^{i\sqrt{4\pi} 
\phi_\sigma(x)}.$$ 
This operator contains bosonic degrees of freedom $\phi_\sigma(x)$, which 
can be integrated out as regular bosonic fields, and a discrete fermion  
number which measures the added changes to the wire, $p_\sigma=0,1,2,...$ 
and $\hat{V}_\sigma$, $\hat{V}^\dag_\sigma$ are the fermionic operators 
(see Eq. (2c)). 

This integration is performed in two stages: (a) Integration of the 
bosonic degrees $\phi_\sigma(x)$.  This step is performed with the 
help of the Renormalization Group.  As a result of this RG the short 
wire Hamiltonian and the tunneling matrix element $\lambda$ are  
renormalized according to the sine-Gordon scaling equations. In addition,  
{\it new magnetic exchange} terms are induced by the RG.  After performing 
the RG we use experimental considerations, namely that the studies of 
conductance have been performed below a few Kelvins for wires with length  
of the order of 1 $\mu$m. This means that the relevant physics occurs at 
a (dimensionless) length scale $l>l_d =\log(d/a)$.  Consequently we compute 
the relevant coupling constants at the scale $l=l_d$.  We substitute 
$\tilde{\Lambda}=1/d$ as our new cutoff and rewrite the model in terms of 
the scaled coupling constant and new fields, 
$$\chi_\sigma(R=\pm (d/2-\epsilon))\rightarrow \hat{V}_\sigma 
e^{i(2\pi/d)(p_\sigma -1/2)R} . \eqno(8) $$ 
Note that in Eq. (8) the bosonic field has completely disappeared.    

The model which is obtained for the short wire corresponds to a {\it 
multilevel} ``quantum dot" system characterized by the quantum number 
$p_\sigma$.  Equation (1a) is replaced by: 
$$\tilde{H}(d)=\tilde{H}^{leads}(d)+\tilde{H}^{wire}(d)+\tilde{H}_T(d) 
+\tilde{H}_{||}(d)+\tilde{H}_{\perp}(d)+\tilde{H}_{p-p}(d)+\tilde{H}_\mu(d) . 
\eqno(9a) $$ 
Here, $\tilde{H}^{leads}(d)$ is the same as $H^{leads}$ except for the fact 
that $\psi_{e,\sigma}\rightarrow\hat{\psi}_{e,\sigma}$ and $\psi_{o,\sigma} 
\rightarrow\hat{\psi}_{o,\sigma}$, where $\hat{\psi}_{e,\sigma}^\dag$, 
$\hat{\psi}_{e,\sigma}$, $\hat{\psi}_{o,\sigma}^\dag$ and 
$\hat{\psi}_{o,\sigma}$ represent the fermions in the leads with the 
cutoff $\tilde{\Lambda}=1/d$ instead of $\Lambda=1/a$. 

$\tilde{H}^{wire}(d)$ represents the renormalized short wire Hamiltonian 
and is given by: 
$$\tilde{H}^{wire}(d)=\frac{\epsilon_0}{2}[p_\uparrow^2+p_\downarrow^2 
+\eta p_\uparrow p_\downarrow]+\hat{V}^\dag_\uparrow\hat{V}_\uparrow 
\hat{V}^\dag_\downarrow\hat{V}_\downarrow[\hat{g}_c(l_d)\delta_{p_\uparrow 
+p_\downarrow,even}+\hat{g}_s(l_d)\delta_{p_\uparrow-p_\downarrow,0}] . 
\eqno(9b)$$ 
In Eq. (9b), $\eta=2U(4\pi)^2/d\epsilon_0\sim1$ is the dimensionless 
interaction parameter; $\hat{g}_c(l_d)$ and $\hat{g}_s(l_d)$ are the 
renormalized Umklapp and backward interactions, respectively, and $p_\sigma$ 
represents the quantum number of the multilevel system.  When 
$(4k_F-G)d\ge2\pi$ we can neglect the Umklapp term.  On the other hand 
the backward interaction term vanishes in the limit $d\rightarrow\infty$. 

The value of $\hat{g}_s(l_d)$ is determined by the sine-Gordon scaling 
equations \cite{neko}.  For a finite $d$ and $K_s>1$ we find, 
$$\hat{g}_s(l_d)\simeq\frac{U}{2\pi^2}\left(\frac{d}{a}\right)^{2(1-K_s)} 
\frac{d}{a^2}=\frac{U}{2\pi^2}\left(\frac{d}{a}\right)^{(3-2K_s)}\frac{1}{d}. 
\eqno(9c) $$ 
Since $K_s(0)>1$ and $d$ is finite, $K_s(l_d)>1$ and therefore the 
critical scaling $d\rightarrow\infty$, $K_s\rightarrow1$ with $\hat{g}_s 
(l\rightarrow\infty)\sim\hat{g}_s(0)/(1+\hat{g}_s(0)l)$ is not applicable. 

We also approximate the Hubbard interaction $U$ by a screened Coulomb 
interaction with a dielectric constant $\kappa=\varepsilon/\varepsilon_0 
\simeq1$.  We approximate $U/a\simeq e^2/(4\pi\varepsilon_0)\kappa a$.  As 
a result we find that $U$ is related to the energy of the short wire 
$\epsilon_0=\hbar v_F/d$, 
$U\simeq\frac{\epsilon_0}{\kappa}\frac{1}{137}(c/v_F)d$, where $c\simeq3 
\times10^8$ m/sec is the speed of light, $c/v_F\simeq10^3$ and $\kappa\simeq 
10$.  Using these values we estimate $K_s(0)=\sqrt{(1+U/\pi t_0v_-)/(1-U/\pi 
t_0v_-)}\simeq\sqrt{(1+(1/137)(c/v_F)(1/\pi\kappa))/(1-(1/137)(c/v_F)(1/\pi\kappa))} 
\sim1.5$.  This estimate allows us to replace Eq. (9c) by $\hat{g}_s(l_d) 
\simeq(\epsilon_0/2\pi^2\kappa)(1/137)(c/v_F)(d/a)^{(3-2K_s)}$.  For $d/a 
\simeq10^3$ we find that $\hat{g}_s(l_d)\sim\epsilon_0$, which is a few meV. 

The next term is the coupling Hamiltonian $\tilde{H}_T(d)$, which replaces 
Eq. (6b) by  
$$\tilde{H}_T(d)=2i\hat{\lambda}\sum_{\sigma=\uparrow,\downarrow}\int_{-L/2}^{L/2} 
dx\delta(x)[\hat{\psi}^\dag_{o,\sigma}(x)\hat{V}_\sigma\cos\left(k_F(d/2- 
\varepsilon)+\frac{2\pi}{d}(p_\sigma-1/2)(d/2-\varepsilon)\right)$$ 
$$-\cos\left(k_F(d/2-\varepsilon)+\frac{2\pi}{d}(p_\sigma-1/2)(d/2- 
\varepsilon)\right)V_\sigma^\dag\hat{\psi}_{o,\sigma}(x)] , \eqno(9d) $$  
where $\hat{\lambda}=(2/\sqrt{\pi d})t\sin(k_Fa)=\tilde{\lambda} 
\tilde{\Lambda}^{1/2}$, $\tilde{\Lambda}=1/d$ and $\tilde{\lambda}=(t/\sqrt{\pi})  
\sin(k_Fa)$.  
This result was obtained within an RG calculation; see Appendix A.      

The next three terms represent the {\it induced} magnetic interaction and 
effective impurity energy:  
$$\tilde{H}_{||}(d)=2\sum_{\sigma=\uparrow,\downarrow}\left[\hat{J}_{||}(l_d) 
+\hat{I}_{||}(l_d)\cos\left(2k_F(d/2-\varepsilon)+\frac{2\pi}{d}(p_\sigma-1/2) 
(d/2-\varepsilon)\right)\right]\hat{\psi}^\dag_{o,\sigma}(0) 
\hat{\psi}_{o,\sigma}(0) \tilde{V}^\dag_\sigma\hat{V}_\sigma .  \eqno(9e) $$ 
The coupling constants $\hat{J}_{||}$ and $\hat{I}_{||}$ are given in 
Appendix A (see Eqs. (A7,A8)).  The transversal exchange term is given by 
$$\tilde{H}_\perp(d)=2\hat{J}_\perp(l_d)\left[\cos\left(\frac{2\pi}{d} 
p_s(d/2-\varepsilon)\right)+\cos\left((2k_F+\frac{2\pi}{d}p_c)(d/2-\varepsilon) 
\right)\right] $$ 
$$\times[\hat{\psi}^\dag_{o,\uparrow}(0) 
\hat{\psi}_{o,\downarrow}\hat{V}^\dag_\downarrow\hat{V}_\uparrow 
+\hat{\psi}^\dag_{o,\downarrow}(0)\hat{\psi}_{o,\uparrow}(0)\hat{V}^\dag_\uparrow 
\hat{V}_\downarrow] , \eqno(9f) $$   
where $p_s=p_\uparrow-p_\downarrow$ and $p_c=p_\uparrow+p_\downarrow$. The 
coupling constant $\hat{J}_\perp(l_d)$ is given in Appendix A (see Eq. (A9)). 
We observe that the coupling Hamiltonian in Eq. (9d) and the induced 
interactions in Eqs. (9e)-(9g) are sensitive to the value of the overlapping 
region $\varepsilon$ and chemical potential $\mu^{wire}$.  The use of 
$\varepsilon\ne0$ and $\mu^{wire}\ne0$ changes the boundary condition 
to $\delta\ne1/2$ [see Eqs. (2a)-(2c)]: 
$(p_\sigma-1/2)(d/2-\varepsilon) =(:p_\sigma:-\delta)(d/2-\varepsilon)$, 
therefore $\delta=1/2-\langle p_\sigma\rangle$. 

The induced two particle term is given by $\tilde{H}_{p-p}(d)$,  
$$\tilde{H}_{p-p}(d)=\hat{J}_{\perp}(l_d)\left[\cos\left(\frac{2\pi}{d}p_s(d/2- 
\varepsilon)\right)+\cos\left((2k_F+\frac{2\pi}{d}p_c) 
(d/2-\varepsilon)\right)\right]$$  
$$[\hat{\psi}^\dag_{o,\uparrow}(0) 
\hat{\psi}^\dag_{o,\downarrow}(0)\hat{V}_\downarrow\hat{V}_\uparrow 
+\hat{V}^\dag_\uparrow\hat{V}^\dag_\downarrow\hat{\psi}_{o,\downarrow}(0) 
\hat{\psi}_{o,\uparrow}(0)] . \eqno(9g)  $$ 
$\tilde{H}_{p-p}(d)$ gives rise to two particle transmission at low 
temperature.  The two particle term dominates when the single 
particle transmission controlled by $\hat{\lambda}$ vanishes.  

Equations (9e) and (9f) represent the induced magnetic interaction. These 
equations are obtained from the first two terms given in Eq. (A6).  The 
particle-particle term given in Eq. (9g) is obtained from the last two 
terms in Eq. (A6).  The explicit form in Eq. (9e) is obtained once we 
integrate out completely the bosonic degrees of freedom $\phi_\sigma(x)$ 
of the short wire.  At scale $l>\log(d/a)$ we replace in Eq. (A6) fermion 
fields $\chi_\sigma$, $\chi^\dag_\sigma$ by the zero mode fermions 
$\hat{V}_\sigma$, $\hat{V}^\dag_{\sigma}$.   

The last term in Eq. (9a) is the induced single particle energy 
$$\tilde{H}_\mu(d)=-\sum_{\sigma=\uparrow,\downarrow}\tilde{E}_\sigma 
\hat{\psi}^\dag_{o,\sigma}(0)\hat{\psi}_{o,\sigma}(0)-\sum_{\sigma= 
\uparrow,\downarrow}E_\sigma(d)\hat{V}^\dag_\sigma\hat{V}_\sigma , \eqno(9h)$$ 
where 
$$E_\sigma(d)=\left(\frac{1}{\pi d}\right)^2\left[\hat{J}_\parallel(l_d) 
+\hat{I}_\parallel(l_d)\cos\left((2k_F+\frac{2\pi}{d}p_\sigma)(d/2-\varepsilon) 
\right)\right]>0 , $$ 
and $\tilde{E}_\sigma\simeq E_\sigma(d)$ are obtained after performing 
the normal ordering in Eq. (A6).  When a gate voltage is applied, $E_\sigma(d)$ 
is shifted by the chemical potential of the wire $\mu^{wire}$ [this will be 
the case in Eq. (15f) below where the single particle energy $\epsilon_1$ 
is shifted, $\epsilon_1\rightarrow \epsilon_1-eV_G$].  The set of Eqs. 
(9a)-(9h) 
complete step one of the renormalization group. 

In the second step we map the problem to an impurity model.  
This step is performed using the impurity Hamiltonian (Eq. 9b). 
We have to project out high energy states: we project out the states 
with $p_\sigma=2,-1,-2,...$.  As a result of this projection we 
keep the same structure as given in Eqs. (9a)-(9h) with the difference 
that $p_\sigma$ is restricted to $p_\sigma=0,1$, $\sigma=\uparrow, 
\downarrow$ and the coupling constants $\hat{J}_\parallel(l_d)$, 
$\hat{I}_\parallel(l_d)$ and $\hat{J}_\perp(l_d)$ are replaced by $\langle 
\hat{J}_\parallel(l_d)\rangle\equiv\overline{J}_\parallel(l_d)$, $\langle 
\hat{I}_\parallel(l_d)\rangle\equiv\overline{I}_\parallel(l_d)$, and 
$\langle\hat{J}_\perp(l_d)\rangle\equiv\overline{J}_\perp(l_d)$ with  
``$\langle...\rangle$" standing for the averages.  From Appendix A we 
find that in the limit $d\rightarrow a$ the induced exchange coupling 
constants $\overline{J}_\parallel(l_d)$, $\overline{I}_\parallel(l_d)$ 
and $\overline{J}_\perp(l_d)$ vanish.  The projected 
Hamiltonian $h$ replaces $\tilde{H}(d)$, $\tilde{H}(d)\rightarrow h$ 
$$h=\tilde{H}^{leads}(d)+h_{imp}+h_T+h_\parallel+h_\perp+h_p , \eqno(10) $$ 
where $\tilde{H}^{wire}(d)+\tilde{H}_\mu (d)\rightarrow h_{imp}$, $\tilde{H}_T(d) 
\rightarrow h_T$, $\tilde{H}_\parallel(d)\rightarrow h_\parallel$,  
$\tilde{H}_\perp(d) \rightarrow h_\perp$, and $\tilde{H}_{p-p}\rightarrow h_p$.  
  
\section{Conductance in the intermediate temperature regime:} 
For $T_{Kondo}\ll T\le \hbar v_F/k_Bd\simeq T^{wire}$  
we use the impurity model obtained in Eq. (9b).  In particular, the 
impurity Hamiltonian with $p_\sigma=0,1$ and $\sigma=\uparrow,\downarrow$ 
is given by, 
$$h_{imp}=\frac{\epsilon_0}{2}[p^2_\uparrow+p^2_\downarrow+\eta p_\uparrow 
p_\downarrow]+\hat{g}_s(l_d)\hat{V}^\dag_\uparrow\hat{V}_\uparrow 
\hat{V}^\dag_\downarrow\hat{V}_\downarrow\delta_{p_\uparrow-p_\downarrow,0} 
-E_d(\hat{V}^\dag_\uparrow\hat{V}_\uparrow+\hat{V}^\dag_\downarrow 
\hat{V}_\downarrow) , \eqno(11a) $$ 
where $\hat{g}_s(l_d)$ is the renormalized backscattering term given in 
Eq. (9c). [We have neglected the Umklapp term $(4k_F-G)d>2\pi$].  In Eq. (9h) 
we replace $E_\sigma(d)$ by $E_d\sim \langle\hat{J}_\parallel\rangle/\pi d$. 

In this temperature regime we ignore the two particle transmission and  
replace $h_\parallel+h_\perp+h_{p-p}$ by $h_\parallel$:  
$$h_\parallel\simeq\langle\hat{J}_\parallel\rangle\sum_{\sigma= 
\uparrow,\downarrow}\hat{\psi}^\dag_{o,\sigma}(0)\hat{\psi}_{o,\sigma}(0) 
\hat{V}^\dag_\sigma\hat{V}_\sigma , \eqno(11b) $$    
with $\langle\hat{J}_\parallel\rangle\simeq(2t^2/\pi v_F)\log(d/a)$. 

The transmission term in Eq. (9d) is replaced by $h_T$.  
$$h_T\simeq i2\hat{\lambda}\sum_{\sigma=\uparrow,\downarrow}\cos(\pi 
p_\sigma)[\hat{\psi}^\dag_{o,\sigma}(0)\hat{V}_\sigma-\hat{V}^\dag_\sigma   
\hat{\psi}_{o,\sigma}(0)] . \eqno(11c) $$ 
We have approximated the cosine term in Eq. (9d) by $\cos(\pi p_\sigma)$. 
This approximation has a negligible effect on the renormalized tunneling  
matrix element $\hat{\lambda}$.  The matrix element $\hat{\lambda}$ is 
fixed by using $\varepsilon\ne0$, $\delta=1/2$ or equivalently $\varepsilon 
=0$, $\delta\ne1/2$, see Eq. (2c).  
 
The impurity spectrum is characterized by the $U(1)\times SU(2)$ 
Kac-Moody primary states of the form $|p_\uparrow,p_\downarrow\rangle$. 
When $p_c=p_\uparrow+p_\downarrow$ is even the state $|p_\uparrow, 
p_\downarrow\rangle=|p_c=even,p_s=0\rangle$ is an $SU(2)$ singlet 
with $p_s=p_\uparrow-p_\downarrow=0$.  On the other hand, when $p_c= 
p_\uparrow+p_\downarrow$ is odd the state $|p_\uparrow,p_\downarrow\rangle 
=|p_c=odd,p_s\pm1\rangle$ is an $SU(2)$ doublet $p_s=\pm1$ and spin 
$p_s/2=\pm1/2$. 

The Heisenberg equations of motion are obtained from the Hamiltonian in 
Eq. (11a) with the zero mode anticommutation relations $\{\hat{V}^\dag_\sigma, 
\hat{V}_{\sigma'}\}=2 \delta_{\sigma,\sigma'}$, $\{\hat{V}_\sigma^\dag, 
\hat{V}_{\sigma'}^\dag\} =\{\hat{V}_\sigma,\hat{V}_{\sigma'}\}=0$ 
and commutation relations $[p_\sigma,\hat{V}_{\sigma'}^\dag] =\delta_{\sigma, 
\sigma'}\hat{V}_{\sigma'}^\dag$, $[p_\sigma,\hat{V}_\sigma]=-\delta_{\sigma, 
\sigma'}\hat{V}_{\sigma'}$. 

We replace the number operator $p_\sigma$ by $\hat{V}^\dag_\sigma 
\hat{V}_\sigma=2\pi(d/a)p_\sigma$.  The factor $2\pi(a/d)$ is related to 
the different renormalization of $p_\sigma$ and $\hat{V}^\dag_\sigma 
\hat{V}_\sigma$.  Next, we determine the eigenvalues for the impurity model 
[Eq. (11a)],   
$$h_{imp}|p_\uparrow,p_\downarrow\rangle=\varepsilon(p_\uparrow,p_\downarrow) 
|p_\uparrow,p_\downarrow\rangle .$$ 
We find that due to the fact that $\hat{g}_s(l_d)\ge\epsilon_0$ and 
$E_d>0$, the $SU(2)$ doublet states $|p_c=odd,p_s=\pm1\rangle$ 
have a lower energy than the singlet state $|p_c=even,p_s=0\rangle$. 
Therefore we expect that the current will be controlled by the $SU(2)$ 
doublet. 

In the remaining part of this section we compute the current 
$I_\sigma$, which is obtained from Eq. (7d) with $H_T$ replaced by 
$h_T$ [see Eq. (11c)], 
$$I_\sigma=\frac{e\hat{\lambda}}{\hbar}\cos(\pi p_\sigma) 
[\hat{\psi}^\dag_{e,\sigma}(0)\hat{V}_\sigma+\hat{V}^\dag_\sigma 
\hat{\psi}_{e,\sigma}(0)] .  \eqno(12a) $$ 
To find the current from Eq. (11) we have to solve for 
the scattering states \cite{david2,simon}.  The scattering states 
will be obtained within the Heisenberg equations of motion for 
$\hat{V}^\dag_\sigma$, $\hat{\psi}_{e,\sigma}$ and $\hat{\psi}_{o,\sigma}$:   
$$i\hbar\dot{\hat{V}}_\uparrow^\dag=[\hat{V}_\uparrow^\dag,h_{imp}+h_T 
+h_\parallel]=-\epsilon_0\left[p_\uparrow-\frac{1}{2}+\frac{\eta}{2}p_\downarrow 
+\frac{1}{2}\overline{g}_s p_\downarrow\delta_{p_\uparrow-p_\downarrow,0}\right] 
\hat{V}_\uparrow^\dag $$ 
$$+E_d\hat{V}^\dag_\uparrow-i(2\hat{\lambda})\cos(\pi p_\uparrow) 
\hat{\psi}_{0,\uparrow}(0)+\langle\hat{J}_\parallel\rangle 
\hat{\psi}^\dag_{0\uparrow}(0)\hat{\psi}_{0\uparrow}(0)\hat{V}_\uparrow . 
\eqno(12b) $$ 
$$i\hbar\dot{\hat{\psi}}_{o\uparrow}(x)=[\hat{\psi}_{o\uparrow}(0),h_{imp} 
+h_\parallel+\tilde{H}^{lead}(d)]=\hbar v_F(-i\partial_x) 
\hat{\psi}_{o,\uparrow}(x) $$ 
$$+\overline{J}_\parallel p_\uparrow\hat{\psi}_{o,\uparrow}(x)\delta(x) 
+i(2\hat{\lambda})\cos(\pi p_\uparrow)\hat{V}^\dag_\uparrow\delta(x) . 
\eqno(12c) $$   
$$i\hbar\dot{\hat{\psi}}_{e\uparrow}(x)=\hbar v_F(-i\partial_x) 
\hat{\psi}_{e,\uparrow}(x) .  \eqno(12d) $$     

The zero mode $p_\sigma$ which measures the charge fluctuations with 
respect to the Fermi energy obeys the Heisenberg equation of motion 
$$i\hbar\dot{p}_\sigma=[p_\sigma,h_{imp}+h_T+h_\parallel]=[p_\sigma,h_T] 
=-2\hat{\lambda}\cos(\pi p_\sigma)[\hat{\psi}_{o,\sigma}^\dag(0)\hat{V}_\sigma 
+\hat{V}_\sigma^\dag\hat{\psi}_{o,\sigma}(0)]. \eqno(12e) $$ 
In Eq. (12b) we have used the notation: $(1/2)\overline{g}_s\equiv 
\hat{g}_s(l_d)2\pi(d/a) (1/\epsilon_0)$ and $\overline{J}_\parallel 
\equiv\langle\hat{J}_\parallel \rangle2\pi(a/d)$.
From Eq. (12c) we find that $\hat{\psi}_{o,\sigma}(0)$ scales with 
$2\hat{\lambda}$ and the spectrum of $\hat{V}_\sigma$ is controlled by the 
single particle energy $\epsilon_0$.  Therefore, we conclude that we can 
neglect the time dependence of $p_\sigma$ when the single particle energy 
is larger than the transmission energy, $(2\hat{\lambda})^2/\epsilon_0<1$. 
In this limit we can neglect the fluctuations $\delta p_\sigma(t)$, 
$p_\sigma(t)=p_\sigma+\delta p_\sigma(t)$ and we make the approximation 
$\langle p_\sigma(t)\rangle\simeq p_\sigma$.  
 
Next we use the bosonic representation of $\hat{\psi}_{o,\sigma}$ and replace, 
$\hat{\psi}_{o\uparrow}^\dag(0) \hat{\psi}_{o,\uparrow}(0)=({1}/{2\pi a}) 
+({1}/{2\pi})\partial_x\theta_{o,\uparrow}$.  Here $\theta_{o,\uparrow}$ is 
the bosonic field in the leads and $\tilde{E}_d$ is the renormalized impurity 
energy, $\tilde{E}_d=E_d+\langle\hat{J}_\parallel\rangle\langle 
\hat{\psi}^\dag_o(0) \hat{\psi}_o(0)\rangle=E_d+\langle\hat{J}_\parallel 
\rangle/2\pi d$.  We neglect the $\partial_x \theta_{o,\uparrow}(0)$ term 
in Eq. (12b) and solve Eqs. (12b) and (12d) together.  We expand 
$\hat{\psi}_{o,\sigma}(x,t)$ in scattering states $U_{E,\sigma}(x)$ with the 
eigenvalues $E$.  Then  
$$\hat{\psi}_{o,\sigma}(x,t)=\sum_E A_{E,\sigma}(t)U_{E,\sigma}(x), \eqno(13a)$$ 
$$\hat{V}_\sigma(t)=\sum_E A_{E,\sigma}(t)\tilde{V}_{\sigma,E} , \eqno(13b)$$ 
with $A_{E,\sigma}(t)=A_{E,\sigma} e^{-i(E/\hbar)t}$.  We find from Eqs. 
(12b), (12d) and the approximation $\langle\delta p_\sigma(t)\rangle=0$ 
that $\tilde{V}^\ast_{\sigma,E}$ is proportional to $U_{E,\sigma}(0)$,  
$$\tilde{V}^\ast_{\uparrow,E}=-\frac{i2\hat{\lambda}\cos(\pi p_\uparrow) 
U_{E,\uparrow}(0)}{\epsilon_0[p_\uparrow-\frac{1}{2}+\frac{\eta}{2}p_\downarrow 
+\frac{\overline{g}_s}{2}p_\downarrow\delta_{p_\uparrow-p_\downarrow,0}] 
-\tilde{E}_d-E} , \eqno(13c) $$ 
and 
$$(E+\hbar v_F(i\partial_x))U_{E,\uparrow}(x)=\overline{J}_\parallel 
p_\uparrow\delta(x)U_{E,\uparrow}(x)+i2\hat{\lambda}\cos(\pi p_\uparrow) 
\tilde{V}^\ast_{\uparrow,E}\delta(x) . \eqno(13d) $$ 
We substitute Eq. (13c) into Eq. (13d) and obtain the scattering state 
solution $U_{E,\uparrow}(x)$:   

$$U_{E,\uparrow}(x)=\frac{1}{\sqrt{L}}e^{i(E/\hbar v_F)x}e^{i\overline{J}_ 
\parallel p_\uparrow\mu(x)}\left[\mu(-x)+\mu(x)\left(\frac{1-i(2\hat{\lambda} 
)^2/S} {1+i(2\hat{\lambda})^2/S}\right)\right] , \eqno(14a) $$ 
where $\mu(x)$ is the step function $\mu(x\ge0)=1$, $\mu(x<0)=0$ and $S$ 
is the resonance energy function 
$$S\equiv S(E;p_\uparrow,p_\downarrow)=\epsilon_0(p_\uparrow-1/2+(\eta/2) 
p_\downarrow+\overline{g}_s p_\downarrow\delta_{p_\uparrow-p_\downarrow,0}) 
-\tilde{E}_d-E , \eqno(14b)$$ 
where $\epsilon_0=h v_F/2d$, $\eta=U(4\pi)^2/\epsilon_0 d$, and 
$\overline{g}_s$ is given by Eq. (9c).  
According to Eq.~(12d), $\hat{\psi}_{e,\sigma}$ is given in terms 
of plane waves, $\hat{\psi}_{e,\sigma}(x)=\sum_E B_{E,\sigma}({1}/{\sqrt{L}})  
e^{i(E/(\hbar v_F))x}$, and $\hat{\psi}_{o,\sigma}(x)=\sum_E A_{E,\sigma} 
U_{E,\sigma}(x)$, where $U_{E,\sigma}(x)$ is the scattering state (Eq. 
(14a)). 

In the leads the presence of the voltage difference $V_{DS}$ gives 
$$\langle(R^\dag_\sigma R_\sigma)_E\rangle=\langle a^\dag_{E,\sigma} 
a_{E,\sigma}\rangle\equiv f_{FD}\left(E+\frac{eV_{DS}}{2}\right) ,$$ 
$$\langle(L^\dag_\sigma L_\sigma)_E\rangle=\langle b^\dag_{E,\sigma} 
b_{E,\sigma}\rangle\equiv f_{FD}\left(E-\frac{eV_{DS}}{2}\right),\eqno(14c) $$ 
where $f_{FD}(E)$ is the thermal Fermi-Dirac function.  The right and 
left moving operators $a_{E,\sigma}$ and $b_{E,\sigma}$ are related 
to $\hat{\psi}_{e,\sigma}$ and $\hat{\psi}_{o,\sigma}$ by the relation: 
$$A_{E,\sigma}=\frac{1}{\sqrt{2}}(a_{E,\sigma}-b_{E,\sigma}); ~~~~ 
B_{E,\sigma}=\frac{1}{\sqrt{2}}(a_{E,\sigma}+b_{E,\sigma}) . \eqno(14d)$$  

We substitute into Eq. (12a) the Eqs. (13a) and (13b) using the eigenfunction 
$U_{E,\sigma}(x)$.  We take the thermal expectation with respect to 
the leads reservoir [see Eq. (7b)] and find:  
$$\langle I_\uparrow\rangle=\frac{e\hat{\lambda}}{\hbar}\cos(\pi p_\uparrow) 
\sum_E\frac{1}{L}\left[\langle B^\dag_{E,\uparrow}A_{E,\uparrow}\rangle 
\tilde{V}_{\uparrow,E}+\langle A^\dag_{E,\uparrow}B_{E,\uparrow}\rangle 
\tilde{V}^\ast_{\uparrow,E}\right] $$ 
$$=-\frac{ie}{4\hbar^2}\sum_E\frac{(2\hat{\lambda})^2}{S(E)}\left 
(U^\dag_{E,\uparrow} 
(0)-U_{E,\uparrow}(0)\right)\left[\langle a^\dag_{E,\uparrow}a_{E,\uparrow} 
\rangle-\langle b^\dag_{E,\uparrow}b_{E,\uparrow}\rangle\right] $$ 
$$=\frac{e}{h}\int dE\left[\frac{(2\hat{\lambda})^4}{(2\hat{\lambda})^4+S^2} 
\cos(\overline{J}_\parallel p_\uparrow)+\frac{(2\hat{\lambda})^2}{S} 
\left(\frac{S^2-(2\hat{\lambda})^4}{S^2+(2\hat{\lambda})^4}\right)  
\sin(\overline{J}_\parallel p_\uparrow)\right] $$ 
$$\times\left[f_{FD}\left(E-\frac{eV_{DS}}{2}\right)-f_{FD}\left(E 
+\frac{eV_{DS}}{2} \right)\right] . \eqno(14e) $$ 
In Eq. (14e) $S$ is given by Eq. (14b).  When we compute the current 
$\langle I_\downarrow\rangle$ we have to replace $p_\uparrow\rightarrow 
p_\downarrow$ and $p_\downarrow\rightarrow p_\uparrow$ in Eq. (14e).  
Next, we perform the expectation value over the impurity states 
at temperature $1/\beta$ using the result in Eq. (14b):  
$$\langle\langle I_\sigma\rangle\rangle=\frac{Tr[e^{-\beta h_{imp}(p_\uparrow, 
p_\downarrow)}\langle I_\sigma\rangle]}{Tr e^{-\beta h_{imp}(p_\uparrow, 
p_\downarrow)}}, ~~~~\sigma=\uparrow,\downarrow .  \eqno(15a) $$ 
Expanding the Fermi-Dirac function in Eq. (14e) around $E=0$ gives  
for the $I_\uparrow$ current the conductance: 
$$G_\uparrow=\frac{\langle\langle I_\uparrow\rangle\rangle}{V_{DS}} 
=\frac{e^2}{h}\left\langle\left\langle\frac{(2\hat{\lambda})^4\cos(\overline{J}_\parallel 
p_\uparrow)}{(2\hat{\lambda})^4+(S(p_\uparrow,p_\downarrow))^2} 
\right\rangle\right\rangle , \eqno(15b) $$ 
where $S(p_\uparrow,p_\downarrow)\equiv S(E=0;p_\uparrow,p_\downarrow)$. 
Here $\langle\langle...\rangle\rangle$ stands for the thermodynamic sum 
over the impurity states: $|p_\uparrow=1,p_\downarrow=0\rangle$, 
$|p_\uparrow=0, p_\downarrow=1\rangle$, $|p_\uparrow=0, p_\downarrow=0 
\rangle$, $|p_\uparrow=1,p_\downarrow=1\rangle$.  Equation (15b) is 
dominated by the resonance term $S(p_\uparrow,p_\downarrow)$ 
which obeys $(2\hat{\lambda})>|S(p_\uparrow,p_\downarrow)|$. 
In obtaining Eq. (15b) we have neglected the term proportional to 
$\sin(\overline{J}_\parallel p_\uparrow)$ in Eq. (14e).  This approximation 
is justified given the fact that $\tilde{J}_\parallel\ll1$, see Eq. (A14). 
In order to perform the expectation value in Eq. (15b) we consider the 
even and odd state eigenvalues $\varepsilon(p_\uparrow,p_\downarrow)$ 
and the resonance term $S(p_\uparrow,p_\downarrow)$.   

{\it The even states}: 
$$\varepsilon(p_\uparrow=0,p_\downarrow=0)=0,~~~S(p_\uparrow=0,p_\downarrow=0) 
=-\left(\frac{\epsilon_0}{2}+\tilde{E}_d\right) ; $$ 
$$\varepsilon(p_\uparrow=1,p_\downarrow=1)=2\epsilon_1+U_{e-e} ,~~~S(p_\uparrow=1, 
p_\downarrow=1)=S(p_\uparrow=0,p_\downarrow=0)+2\epsilon_1+U_{e-e} . \eqno(15c) $$ 
   
{\it The odd states}: 
$$\varepsilon(p_\uparrow=1,p_\downarrow=0)=\epsilon_1 ,  
~~~S(p_\uparrow=1,p_\downarrow=0) = \epsilon_1 ; $$ 
$$\varepsilon(p_\uparrow=0,p_\downarrow=1)= \epsilon_1 ,  
~~~S(p_\uparrow=0,p_\downarrow=1)=\epsilon_1+\left(\frac{\eta}{2}-1\right) 
\epsilon_0 , \eqno(15d) $$ 
where $\epsilon_1\equiv(\epsilon_0/2-\tilde{E}_d$) is the ``impurity" single 
particle energy which is below the Fermi energy. $U_{e-e}\equiv\epsilon_0(\eta/2 
+\overline{g}_s)$ is the effective ``Hubbard" interaction and $S(p_\uparrow=0, 
p_\downarrow=1)-S(p_\uparrow=1,p_\downarrow=0)\equiv(\eta/2-1)\epsilon_0$ is 
the effective spin polarization energy.  The energies 
$\varepsilon(p_\uparrow,p_\downarrow)$ control the thermal expectation 
values.  We observe that due to the electron-electron interaction 
$\overline{g}_s$ and the single particle energy $\tilde{E}_d$ the odd states, 
$\varepsilon(p_c=odd,p_s=\mp1)$, can have lower energy than the even states.  

The denominator function $S(p_\uparrow,p_\downarrow)$ controls the quantum 
weights for the different states $|p_\uparrow,p_\downarrow\rangle$ and it  
is asymmetric under the transformation $p_\uparrow\rightarrow p_\downarrow$, 
$p_\downarrow\rightarrow p_\uparrow$.  For the odd states it has the property 
for $I_\uparrow$, $S(p_\uparrow=0,p_\downarrow=1)\ne S(p_\uparrow=1, 
p_\downarrow=0)$. [Similarly, for $I_\downarrow$, $S(p_\uparrow=0, 
p_\downarrow=1) \ne S(p_\uparrow=1,p_\downarrow=0)$]. 
Due to these properties we note that the current is dominated by one of 
the two states.  We use Eqs. (15b)-(15d) to find:  
$$G_\uparrow=\frac{e^2}{h}[\frac{(2\hat{\lambda})^4}{(2\hat{\lambda})^4 
+(S(0,0))^2}+\frac{(2\hat{\lambda})^4\cos(\overline{J}_\parallel)}{(2 
\hat{\lambda})^4 +(S(1,1))^2} e^{-\beta\varepsilon(1,1)}+\frac{(2 
\hat{\lambda})^4\cos(\overline{J}_\parallel)} {(2\hat{\lambda})^4 
+(S(1,0))^2}e^{-\beta \varepsilon(1,0)} $$ 
$$+\frac{(2\hat{\lambda})^4}{(2\hat{\lambda})^4+(S(0,1))^2}e^{-\beta 
\varepsilon(0,1)}] [1+e^{-\beta\varepsilon(1,1)}+e^{-\beta\varepsilon(1,0)} 
+e^{-\beta\varepsilon(0,1)}]^{-1} . \eqno(15e) $$

The conductance in Eq. (15a) depends on the effective single ``particle" 
energy, $\epsilon_1\equiv\epsilon_0/2-\tilde{E}_d<0$, the vacuum energy, 
$\epsilon_c\equiv \epsilon_0/2+\tilde{E}_d\equiv-S(p_\uparrow=0, 
p_\downarrow=0)$, the effective two particle energy $U_{ee}=\epsilon_0 
(\eta/2+\overline{g}_s)>0$ and polarization energy $\epsilon_s\equiv 
\epsilon_0(\eta/2-1)$.  In the presence of the gate voltage $V_G$ the 
single particle energy is shifted $\epsilon_1\rightarrow\epsilon_1-eV_G$. 
(The single particle energy $\epsilon_1$ is controlled by the renormalized 
wire energy and by the chemical potential shift $\mu^{wire}=eV_G$.)  

Next we introduce dimensionless parameters: $(2\hat{\lambda})^2/\epsilon_0 
\equiv\Gamma$, $\epsilon_1/\epsilon_0\equiv\hat{\epsilon}_1$, $eV_G/ 
\epsilon_0\equiv\hat{U}_G$, $U_{ee}/\epsilon_0\equiv\hat{U}_{ee}$, 
$\epsilon_c/\epsilon_0\equiv\hat{\epsilon}_c$, $\epsilon_s/\epsilon_0 
\equiv\hat{\epsilon}_s$, $\epsilon_0/k_BT\equiv T^{wire}/T$, and 
$\cos(\overline{J}_\parallel)\simeq1$.  
As a consequence we find that the conductance takes the form: 
$$G_\uparrow=\frac{e^2}{h}\frac{1}{Z}[\frac{\Gamma^2}{\Gamma^2+ 
(\hat{\epsilon}_c+\hat{U}_G)^2}+e^{-\hat{\beta}[2(\hat{\epsilon}_1-\hat{U}_G) 
+(\eta/2+\overline{g}_s)]}\frac{\Gamma^2}{\Gamma^2+[2(\hat{\epsilon}_1 
-\hat{U}_G)+(\eta/2+\overline{g}_s)]^2}$$ 
$$+e^{-\hat{\beta}(\hat{\epsilon}_1 
-\hat{U}_G)}\left(\frac{\Gamma^2}{\Gamma^2+(\hat{\epsilon}_1-\hat{U}_G)^2} 
+\frac{\Gamma^2}{\Gamma^2+[(\hat{\epsilon}_1-\hat{U}_G)+(\eta/2-1)]^2} 
\right)] ,  \eqno(15f)$$ 
where $Z=1+e^{-\hat{\beta}[2(\hat{\epsilon}_1-\hat{U}_G)+(\eta/2 
+\overline{g}_s)]}+2e^{-\hat{\beta}(\hat{\epsilon}_1-\hat{U}_G)}$.  
Due to the repulsive interactions $2\hat{\epsilon}_1+\hat{U}_{ee}>0$ 
and $\hat{\epsilon}_1<0$, we find that for $T^{wire}>T$ the even part 
$p_c=0$, 2 in Eq. (15f) can be neglected.  Consequently for $\hat{U}_G=0$ 
Eq. (15f) is replaced by: 
$$G_\uparrow\simeq\frac{e^2}{h}\frac{1}{2+e^{(T^{wire}/T)\hat{\epsilon}_1}} 
\left[\frac{\Gamma^2}{\Gamma^2+\hat{\epsilon}_1^2} +\frac{\Gamma^2} 
{\Gamma^2+(\hat{\epsilon}_1+\eta/2-1)^2}\right]. \eqno(15g)$$   
 
We investigate Eq. (15g) for $\hat{\epsilon}_1<0$.  As a result the term 
$e^{(T^{wire}/T)\hat{\epsilon}_1}\rightarrow0$ when $T^{wire}/T\gg1$.  Due  
to the repulsive interaction $\eta/2$ we have a situation where one 
of the two terms satisfies the condition $(\hat{\epsilon}_1/\Gamma)^2<1$ 
or $(\hat{\epsilon}_1+\eta/2-1)^2/\Gamma^2<1$.  Choosing $\hat{\epsilon}_1 
+\eta/2-1=0$ we find: 
$$G_\uparrow\simeq\frac{e^2}{2h}\left[1+\frac{\Gamma^2}{\Gamma^2 
+(\eta/2-1)^2} \right]\simeq\frac{e^2}{2h}. \eqno(15h) $$


A similar calculation for $G_\downarrow$ gives $G_\downarrow\simeq 
\frac{e^2}{2h}$.  Therefore $G=G_\uparrow+G_\downarrow\simeq\frac{e^2}{h}$. 
This result is not based on a spin polarized state but makes use of the 
asymmetry betweed odd and even states such that only one state gives a 
resonant contribution $(2\hat{\lambda})^4/((2\hat{\lambda})^4+S^2)\sim1$ 
for $S(0,1)$ or $S(1,0)$ but not both.   

The result obtained above depends on the number of electrons in the 
wire.  We consider the case $p_\uparrow+p_\downarrow=p_c=1$. (The even 
case $p_c=0,2,...$ can be ignored since these correspond to high 
energy states and therefore give negligible contribution to the 
current).  When $p_c=1$ we have one electron in the highest state in 
the wire.  Suppose that an additional electron transmits into the 
wire.  Here we can have the following two situations: (a) The spin 
of the electron in the wire is the {\it same} as the spin of the 
transmitting electron;  (b) The spin of the electron in the wire is 
opposite to the one which is transmitting into the wire.  For case 
(a) the transmitting electron having the same spin as the one in the 
wire must occupy the next level and therefore the energy is increased 
by $\epsilon_0$ (by the level separation).  For the case (b) when the 
spins are opposite, the transmitting electron can occupy the same level 
with the electron in the wire.  As a consequence, the energy of the 
system will increase only by $\epsilon_0\eta$ (due to the repulsive 
interaction in the wire). 
This is the reason for the asymmetric resonant condition $S(p_\uparrow=1, 
p_\downarrow=0)\ne S(p_\uparrow=0, p_\downarrow=1)$ given in Eq. (15f) 
and the single particle asymmetry in energy, $\epsilon(p_\uparrow-1/2+ 
(\eta/2)p_\downarrow)$ [see Eq. (12b)]. This formula is asymmetric under 
the transformation $p_\uparrow \rightarrow p_\downarrow$ and 
$p_\downarrow\rightarrow p_\uparrow$ for $p_c=1$.  As a result 
the current is dominated by electrons which have opposite spins to 
the one in the wire.  It might appear from this result that the transmission  
current is polarized. Since the electron in the short wire is in contact 
with a ``thermal bath" the electron has equal probability 
to be in a state $|p_\uparrow=0, p_\downarrow=1\rangle$ or $|p_\uparrow=1, 
p_\downarrow=0\rangle$.  Due to this fact we conclude that the current 
is {\it not} polarized since for each polarized incoming electron only 
one of the two wire {\it states} $|p_\uparrow=0, p_\downarrow=1\rangle$ 
or $|p_\uparrow=1, p_\downarrow=0\rangle$ gives rise to a transmission  
current.  We conclude that the conductance for each polarization is 
half the value of the noninteracting conductance $G_\uparrow\sim 
G_\downarrow\sim e^2/2h$.  Consequently we find that $G=G_\uparrow+ 
G_\downarrow\sim e^2/h$.   


We remark that Eq. (15g) gives perfect conductance when both 
$\hat{\epsilon}_1^2$ and $(\hat{\epsilon}_1+(\eta/2-1))^2$ obey the 
condition $\hat{\epsilon}_1^2/\Gamma^2<1$, $(\hat{\epsilon}_1+(\eta/2-1) 
)^2/\Gamma^2<1$.  Under this condition we obtain from Eq. (15g) $G_\uparrow 
\simeq e^2/h$, and consequently $G=G_\uparrow+G_\downarrow\simeq2e^2/h$. 
The result obtained for this case is in a region where our calculation 
in Eqs. (12b)-(14b) might not be valid since $\Gamma^2$ is large.  For a 
short wire $\overline{g}_s$ is large and Eq. (12e) in conjunction with 
the subsequent discussion can invalidate the condition $\dot{p}_\sigma 
\simeq0$.  This case is investigated explicitly in the next section.  
 
\section{Conductance in the low temperature regime} 
Equation (10) represents our effective impurity model at length scale 
$l>l_d$.  At this length scale the Hamiltonian impurity model $h_{imp}$  
[see Eq. (11a)] contains an effective Hubbard interaction, $\hat{g}_s(l_d) 
\hat{V}^\dag_\uparrow\hat{V}_\uparrow\hat{V}^\dag_\downarrow\hat{V}_\downarrow$. 
For finite $l_d$ we can have a situation where the tunneling matrix element 
$\hat{\lambda}$ in the coupling Hamiltonian $h_T$ [see Eq. (11c)] obeys  
$\hat{\lambda}<\hat{g}_s(l_d)$.  [$\hat{\lambda}\propto d^{-1/2}$, $\hat{g}_s 
\sim(d/a)^{3-2K_s)}(1/d)$].  Here we investigate this case, namely 
$\hat{\lambda}/\hat{g}_s<1$.  This situation is easily achieved in a short 
wire where $\hat{g}_s(l_d)$ is close to the bare Hubbard interaction $U$,  
which obeys $t/U\ll1$.  (For a long wire $\hat{g}_s$ decreases, first 
exponentially and then logarithmically, contrary to $\hat{\lambda}$ which 
decreases like $d^{-1/2}$.  Therefore, for a long wire the condition 
$\hat{\lambda}/\hat{g}_s<1$ is {\it not achieved}).  In the limit 
$\hat{\lambda}/\hat{g}_s<1$ we have a strong coupling problem. 

The spin degrees of freedom of the impurity interact with the conduction 
electrons in the leads.  A virtual charge fluctuation in which an electron 
migrates off or onto the impurity gives rise to a spin-exchange between the 
impurity and the electrons in the leads.  As a result an antiferromagnetic 
interaction between the impurity and the electron in the leads is induced 
\cite{david3,coleman}.  Following Eq. (10), we consider only the impurity 
model plus transmission plus the chemical potential part given in Eqs. (9d) 
and (9h):   
$$\overline{h}=h_{imp}+h_T+h_\mu=\frac{\epsilon_0}{2}[p_\uparrow+p_\downarrow 
+\eta p_\uparrow p_\downarrow]-\tilde{E}_d[\hat{\psi}^\dag_{o,\uparrow}(0)   
\hat{\psi}_{o,\uparrow}(0)+\hat{\psi}^\dag_{o,\downarrow}(0)\hat 
{\psi}_{o,\downarrow}(0)]  $$ 
$$+\hat{g}_s(l_d)\hat{V}^\dag_\uparrow\hat{V}_\uparrow\hat{V}^\dag_\downarrow 
\hat{V}_\downarrow+i2\hat{\lambda}\sum_{\sigma=\uparrow,\downarrow}\cos 
(\pi p_\sigma)[\hat{\psi}_{o,\sigma}^\dag(0)\hat{V}_\sigma-\hat{V}_\sigma 
^\dag\hat{\psi}_{0,\sigma}(0)] ,  \eqno(16) $$  
where $p_\sigma=0,1$, $\sigma=\uparrow,\downarrow$ and $\tilde{E}_d\ll E_d$ 
with $E_d>0$.  

Using the projection method we project out the double occupancy for 
$\hat{V}^\dag_\sigma$, $\hat{V}_\sigma$ and $\hat{\psi}^\dag_{o,\sigma}(0)$,  
$\hat{\psi}_{o,\sigma}(0)$:  
$$\hat{V}^\dag_\sigma=|0\rangle\langle\sigma|+e_{\sigma,-\sigma} 
|-\sigma\rangle\langle\uparrow,\downarrow| , $$ 
$$\hat{V}_\sigma=|\sigma\rangle\langle0|+e_{\sigma,-\sigma} 
|\uparrow,\downarrow\rangle\langle-\sigma| , \eqno(17a) $$ 
and 
$$\hat{\psi}^\dag_{o,\sigma}=|0)(\sigma|+e_{\sigma,-\sigma} 
|-\sigma)(\uparrow,\downarrow| , $$ 
$$\hat{\psi}_{o,\sigma}=|\sigma)(0|+e_{\sigma,-\sigma} 
|\uparrow,\downarrow)(-\sigma| , \eqno(17b) $$ 
where $e_{\uparrow,\downarrow}=1=-e_{\downarrow,\uparrow}$. 
    
The full Hilbert space consists of $1_V\otimes1_{\hat{\psi}(0)}\equiv1$ 
$$1_V=|0\rangle\langle0|+|\uparrow\rangle\langle\uparrow|+|\downarrow\rangle 
\langle\downarrow|+|\uparrow,\downarrow\rangle\langle\downarrow,\uparrow| ,$$ 
$$1_{\hat{\psi}(0)}=|0)(0|+|\uparrow)(\uparrow|+|\downarrow)(\downarrow| 
+|\uparrow,\downarrow)(\downarrow,\uparrow| .  \eqno(17c) $$   
We project out the double occupancy 
$$Q=Q_V\otimes Q_{\hat{\psi}(0)}=(|\uparrow,\downarrow\rangle\langle\downarrow, 
\uparrow|)\otimes(|\uparrow,\downarrow)(\downarrow,\uparrow|) . 
\eqno(17d) $$ 
As a result we obtain, in agreement with Refs. \cite{david3,coleman}, 
$$h_{eff}=-P\overline{h}Q(Q\overline{h}Q-E)^{-1}Q\overline{h}P 
=J_{eff}\vec{s}(0)\cdot\vec{S} , \eqno(18) $$ 
where $P=1-Q$ and 
$$\vec{s}(0)=\hat{\psi}^\dag_{o,\alpha}(0)\left(\frac{\vec{\sigma}}{2}\right 
)_{\alpha,\beta} \hat{\psi}_{o,\beta}(0) , ~~~~\vec{S}=\hat{V}^\dag_\alpha 
\left(\frac{\vec{\sigma}}{2}\right)_{\alpha,\beta}\hat{V}_\beta ; $$ 
$$J_{eff}=2(2\hat{\lambda})^2\left[\frac{1}{\hat{g}_s(l_d)-E_d}+\frac{1}{E_d} 
\right] .  \eqno(19) $$ 

For a short wire such that $\hat{\lambda}/\hat{g}_s(l_d)<1$ we can neglect 
the induced terms $h_\parallel$ and $h_\perp$.  (For a long 
wire this  condition is not possible to satisfy.)  Therefore, for this case 
we rely on the result given in Eqs. (14e), (15b) and (15e) for single particle 
transmission. For the remaining discussion we consider the strong coupling 
limit.  

In the strong coupling limit we consider the Hamiltonian obtained in Eq. (18), 
$$H_K=H^{leads}+J_{eff}\vec{S}\cdot\vec{s}(0) .  \eqno(20a) $$ 
The RG equation for the Hamiltonian in Eq. (20a) are \cite{david3,coleman}  
$$\frac{dJ_{\parallel,eff}}{dl}=\frac{1}{\pi v_F}J^2_{\perp,eff} , ~~~~ 
\frac{dJ_{\perp,eff}}{dl}=\frac{1}{\pi v_F}J_{\parallel,eff}J_{\perp,eff} . 
\eqno(20b) $$ 
For $J_{\parallel,eff}=J_{\perp,eff}=J_{eff}$ we find \cite{david3,coleman} 
that $J_{eff}$ flows to the strong coupling value $J_{eff}(l)=\frac{J_{eff}} 
{1-2J_{eff}l}$.  This allows us to introduce the Kondo temperature $T_K$, 
$$T_K=\frac{\hbar v_F}{k_B}\exp\left[-\frac{1}{2J_{eff}}\right]. \eqno(20c) $$ 
   
At temperature $T<T_K$ the symmetric electron ``orbit" is screened out by the 
impurity.  The symmetric screened electron state is given by $c_{s,\sigma}$:  
$$c_{s,\sigma}=\frac{1}{\sqrt{2}}[c_{L,\sigma}(-d/2-\varepsilon)+c_{R,\sigma}(d/2 
+\varepsilon)]=2i(\sin k_F\varepsilon)\hat{\psi}_{o,\sigma}(0) . \eqno(20d) $$ 
Here $c_{L,\sigma}$ and $c_{R,\sigma}$ are the electrons in the leads.  Since 
the symmetric orbital $\hat{\psi}_{o,\sigma}(0)$ is screened away we 
project it out into the asymmetric orbital 
$$c_{a,\sigma}=\frac{1}{\sqrt{2}}[c_{L,\sigma}(-d/2-\varepsilon)-c_{R,\sigma}(d/2 
+\varepsilon)]=2i(\sin k_F\varepsilon)\hat{\psi}_{e,\sigma}(0) . \eqno(20e) $$ 
Using the projection operator $P=|\hat{\psi}_{e,\sigma}\rangle\langle\hat 
{\psi}_{e,\sigma}|$ we replace $R_\sigma$ and $L_\sigma$ by   
$PR_\sigma P=\frac{1}{\sqrt{2}}\hat{\psi}_{e,\sigma}$ and $PL_\sigma 
P=\frac{1}{\sqrt{2}}\hat{\psi}_{e,\sigma}$.  We substitute these projections 
into the boundary term $H_{BC}$ given in Eq. (1g) and $H^{leads}$ given by 
Eq. (1h).  Thus we obtain  
$$H_e=\sum_{\sigma=\uparrow,\downarrow}{\hbar v_F}\int_{-L/2}^{L/2} 
dx[\hat{\psi}^\dag_{e,\sigma}(x)(-i\partial_x)\hat{\psi}_{e,\sigma}(x) 
-\delta(x)(2\sin k_F a)^2\hat{\psi}^\dag_{e,\sigma}(x)\hat{\psi}_{e,\sigma}(x)] . 
\eqno(21) $$  
Here $H_e$ represents the effective Kondo screened Hamiltonian expressed in 
terms of the asymmetric orbitals $\hat{\psi}_{e,\sigma}$.  For $k_Fa=\pi/2$ 
Eq. (21) represents the single impurity resonant level model.  This model 
has {\it perfect transmission}, namely $G_\uparrow= G_\downarrow=e^2/h$, 
and $G=2e^2/h$.  

In summary, we have obtained perfect transmission, $G=2e^2/h$, in the region 
$T<T_K$ and for a short wire such that $\hat{\lambda}/g_s(l_d)<1$.  This is in  
contrast to a long wire and intermediate temperatures, where $G\simeq e^2/h$.  
At low temperatures and a long wire the conductance is affected by the 
{\it two particle} transmission, see Eq. (9g). 

\section{Discussion} 
To compare our theory with the existing experiments 
\cite{thomas,reilly,reilly1} we used Eq. (15f) to compute the 
conductance.  This equation was obtained under the assumption that 
$\dot{p}_\sigma\simeq0$. This assumption is consistent with the assumption 
of weak transmission; therefore in the Kondo regime our formula might not 
be valid.  However, we find that Eq. (15f) works rather well and explains 
both of the regimes of a long wire with weak transmission and a short wire 
with large transmission. Using Eq. (15f) we observe that the shape of the 
conductance curve is mainly dependent on the tunneling matrix element 
and on the gate voltage $\hat{U}_G\equiv eV_G/\epsilon_0$.  We plot 
the conductance $G/(2e^2/h)$, $G=G_\uparrow+G_\downarrow$ as a 
function of $\hat{U}_G$, for different $\Gamma^2\equiv(2\hat{\lambda})^2/ 
\epsilon_0$ tunneling matrix elements and different $\hat{\beta}= 
\epsilon_0/k_BT=T^{wire}/T$ temperatures. 

The conductance is less sensitive to the rest of the parameters $\eta/2$, 
$\overline{g}_s$, electron-electron interaction and single particle 
energy.  We use $\overline{g}_s\simeq\eta/2\simeq0.75$, $\hat{\epsilon}_1 
=-0.2$ and $\hat{\epsilon}_c=1.2$ [see Eq. (15f)]. In Fig. 1 we plot the 
calculated conductance as a function of gate voltage $\hat{U}_G$ for 
fixed transmission $\Gamma^2=0.1$ (which corresponds to weak transmission) and 
temperatures $\hat{\beta}=1$, 10 and 50.  At $\beta=1$ we observe that the 
conductance takes the value of $G/(2e^2/h)\simeq0.8$.  Decreasing 
the temperature to $\hat{\beta}=10$ we observe that the conductance has 
become 1 and in addition we observe a shoulder in the conductance around 
0.7.  At $\beta=50$, the shoulder develops into a separate conductance 
peak.  Note that this figure very much resembles the experimentally 
measured conductance \cite{thomas,reilly,reilly1}.  

To understand the situation for large tunneling matrix elements 
(i.e. a short wire and in the Kondo regime) we chose $\Gamma^2 =5$ 
and $\hat{\beta}=100$.  In Fig. 2 we plot the calculated conductance 
as a function of gate voltage for these parameters.  We observe 
that the conductance is $G/2e^2/h=1$ as is expected from the 
Kondo solution given in Sec. V.  However, we note that in this 
regime the approximation used in Eq. (15f) is not valid and the 
Kondo solution given in Sec. V should be used.      

\section{Conclusion}  
We have investigated the case in which a two dimensional electronic 
waveguide with a varying width $D(x)$ in the $y$ direction and $e$-$e$ 
interactions can be projected to a one channel problem.  In this case 
the reservoirs are replaced by a one dimensional Fermi liquid (leads) 
coupled to a short wire Luttinger liquid controlled by the gate 
voltage $\mu^{wire}$. Using a combined 
method of Renormalization Group and zero mode bosonization, we have 
constructed the effective spectrum of a short interacting wire.  This 
spectrum consists of combined charge-spin density waves and is dominated 
by the zero mode fermionic spectrum.  At temperature $T\le T^{wire}$ the 
fermionic spectrum is equivalent to a multilevel spectrum for which we 
find the conductance formula given in Eq. (15b).  Transmission is dominated 
by the spin $S=\pm1/2$ doublet giving rise to a conductance $G\sim e^2/h$.  
This result is relevant for experiments in which the wire is short, $T\le 
T^{wire}\simeq\hbar v_F/k_Bd$.  On the other hand, at lower temperatures 
when the effective tunneling matrix $\hat{\lambda}$ is smaller than the 
effective electron-electron interaction $g_s(l_d)$, $\hat{\lambda}/g_s(l_d) 
<1$, we find that the short interacting wire is mapped into a Kondo problem 
which gives rise, for $T<T_{Kondo}< T^{wire}$, to a conductance $G=2e^2/h$.  
By tuning the length of the wire and the temperature we obtain $e^2/h\le 
G\le 2e^2/h$ in agreement with experiments \cite{reilly}.  

\section{Acknowledgment} 
This work was supported by the Los Alamos Laboratory Directed
Research and Development Program.  D.S. was supported by the U.S.
Department of Energy grant number FG02-01ER45909.

\section{appendix A} 
Here, using the RG equation we derive the coupling constants for the 
model in Eq. (9a).  (We use the scaling results for the short wire.)  We 
treat, within a perturbative RG, the transmission term $H_T$. We separate 
$\phi_\sigma(x)=\phi^<_\sigma(x)+\delta\phi_\sigma(x)$, $\psi_{o,\sigma}(x) 
=\psi^<_{o,\sigma}(x)+\delta\psi_{o,\sigma}(x)$, $\psi_{e,\sigma}(x)= 
\psi^<_{e,\sigma}(x)+\delta\psi_{e,\sigma}(x)$ and integrate $\delta\phi_ 
\sigma(x)$, $\delta\psi_{o,\sigma}(x)$, and $\delta\psi_{e,\sigma}(x)$ 
reducing the cutoff from $\Lambda$ to $\Lambda'=\Lambda-d\Lambda\equiv 
\Lambda e^{-l}$.  As a result of the integration we replace $H_T$ 
with $H_{T,eff}=H_T^<+dH_T^{(1)}+dH_T^{(2)}$, where $H^<_T$ represents the 
transmission term at the reduced cutoff and fields $\phi^<_\sigma(x)$, 
$\psi^<_{o,\sigma}(x)$, $\psi^<_{e,\sigma}(x)$, and $dH_T^{(1)}$ 
represents the first order correction in $\lambda$:    
$$dH_T^{(1)}=-i\lambda\frac{\sqrt{4\pi}}{2}[\langle\delta\theta_{o,\uparrow}^2 
\rangle+\langle\delta\theta_{o,\downarrow}^2\rangle+\langle\delta\phi^2_\uparrow 
\rangle+\langle\delta\phi^2_\downarrow\rangle]\sum_{R=\pm(d/2-\varepsilon)} 
\int_{L/2}^{L/2}dx\delta(x)[\psi^{\dag<}_{o,\sigma}(x)e^{ik_FR}\chi^<_\sigma(R)$$  
$$-e^{-ik_FR}\chi^{\dag<}_\sigma(x)\psi^<_{o,\sigma}(x)] . \eqno(A1) $$ 
In Eq. (A1) we have used the bosonic representation of the fermions in the leads, 
$\psi_{o,\sigma}(x)=\frac{1}{\sqrt{2\pi a}}e^{i\sqrt{4\pi}\theta_{o,\sigma}(x)}$ 
and the bosonic part $(1/\sqrt{2\pi a})\exp(i\sqrt{4\pi}\phi_\sigma(x))$ of the 
short wire, see Eq. (2c).  We obtain 
$$\sum_{\sigma=\uparrow,\downarrow}\langle\delta\theta^2_{o,\sigma}\rangle 
=\frac{2\pi}{8}dl , \eqno(A2) $$ 
$$\sum_{\sigma=\uparrow,\downarrow}\langle\delta\phi_\sigma^2\rangle=\frac 
{K_c+K_s}{8\pi}dl , \eqno(A3) $$ 
where $K_c$ and $K_s$ are scale dependent (see \cite{reilly1}).  Using Eqs. 
(A2) and (A3) we find 
$$\frac{d\lambda}{dl}=-\left(\frac{2+K_c+K_s}{4}\right)\lambda . \eqno(A4) $$ 
  
To a good approximation $K_c+K_s\simeq2$.  This gives $d\lambda/dl\simeq 
\lambda$ with the scaling law $\lambda(l)=\lambda e^{-l}$.  Therefore 
$\lambda(l_d)=\lambda(d/a)^{-1}$.  Next, we rescale the fields such that 
$\psi_{o,\sigma}(x)$ is replaced by $\hat{\psi}_{o,\sigma}(x)$, $\psi_{o,\sigma}(x) 
=\sqrt{d/a}\hat{\psi}_{o,\sigma}(x)$, where $\hat{\psi}_{o,\sigma}(x)$ 
depends on the new cutoff $\tilde{\Lambda}=1/d$.  Since at the scale $l>l_d$ 
the bosonic field has disappeared, we replace $V_\sigma=1/\sqrt{2\pi a} 
\hat{V}_\sigma$ by $\hat{V}_\sigma$.  As a result the new coupling constant 
becomes, 
$$\hat{\lambda}\equiv\lambda(l_d)\left(\frac{d}{a}\right)^{1/2}\frac{1} 
{\sqrt{2\pi a}}\equiv\tilde{\lambda}\tilde{\Lambda}^{1/2}, ~~~~ \tilde{\lambda} 
=\frac{t}{\sqrt{\pi}}\sin k_Fa, ~~~~ \tilde{\Lambda}\equiv1/d . \eqno(A5) $$ 

The induced RG terms are obtained from the second order term in $\lambda$, 
$dH_T^{(2)}=dH_{T,p-h}^{(2)}+dH_{T,p-p}^{(2)}$, where $dH_{T,p-h}^{(2)}$ 
represents the particle-hole and $dH_{T,p-p}^{(2)}$ the two-particle 
induced interactions:   
$$dH_T^{(2)}=\frac{i}{2}(i\lambda)^2\sum_{\sigma,\sigma'=\uparrow,\downarrow} 
\sum_{R,R'=\pm(d/2-\varepsilon)}\int dt\int d\tau T\{[\psi^{\dag <} 
_{o,\sigma}(0,t+\tau)e^{ik_FR}\chi_\sigma^<(R,t+\tau) e^{-ik_FR'} 
\chi_{\sigma'}^{\dag <} (R',t)$$ 
$$\psi^<_{o,\sigma'}(0,t)+e^{-ik_FR}\chi^{\dag <}_\sigma(R,t+\tau) 
\psi_{o,\sigma}^<(0,t+\tau)\psi_{o,\sigma'}^{\dag <}(0,t)e^{ik_FR'} 
\chi_{\sigma'}^<(R',t)]~[4\pi\langle\delta\theta_{o,\sigma}(t+\tau)\delta 
\theta_{o,\sigma'}(t) \rangle $$ 
$$+4\pi\langle\delta\phi_\sigma(R,t+\tau)\delta\phi_{\sigma'}(R',t)\rangle]$$ 
$$+[\psi_{o,\sigma}^{\dag <}(0,t+\tau)e^{ik_FR}\chi_\sigma^<(R,t+\tau)e^{ik_FR'} 
\chi_{\sigma'}^<(R',t+\tau)\psi_{o,\sigma'}^{\dag <}(0,t)$$ 
$$+e^{-ik_FR}\chi_\sigma^{\dag <}(R,t+\tau)\psi_{o,\sigma}^<(0,t+\tau) 
\psi_{o,\sigma'}^<(0,t+\tau)e^{-ik_FR'}\chi_{\sigma'}^{\dag <}(R',t)][4\pi 
\langle\delta\theta_{o,\sigma}(t+\tau)\delta\theta_{o,\sigma'}(t)\rangle $$ 
$$+4\pi\langle\delta\phi_\sigma(t+\tau)\delta\phi_{\sigma'}(t)\rangle]\}.\eqno(A6)$$ 
Using the short wire RG equation we compute the expectation values of 
$\langle\delta\phi_\sigma\delta\phi_{\sigma'}\rangle$.  Consequently, we 
obtain the RG equations for $J_\parallel$ and $I_\parallel$ which appear in 
Eqs. (9e)-(9g):  
$$\frac{dJ_\parallel}{dl}=-J_\parallel+ \frac{\lambda^2}{v_F\Lambda(l)} 
\{1+\frac{v_F}{2}[\frac{(K_c(l)v_c(l))^{-1}+(K_s(l)v_s(l))^{-1}}{2}(1+\cos 
\Lambda(l)R)$$ 
$$+\frac{(K_c(l)/v_c(l))+(K_s(l)/v_s(l))}{2}(1-\cos\Lambda(l)R)]\}. \eqno(A7) $$ 
$$\frac{dI_\parallel}{dl}=-I_\parallel+ \frac{\lambda^2}{v_F\Lambda(l)} 
\{1+\frac{v_F}{2}[\frac{(K_c(l)v_c(l))^{-1}+(K_s(l)v_s(l))^{-1}}{2}(1+\cos
\Lambda(l)R)$$
$$-\frac{(K_c(l)/v_c(l))+(K_s(l)/v_s(l))}{2}(1-\cos\Lambda(l)R)]\}. \eqno(A8) $$  
The RG equation for the transversal exchange coupling constant $J_\perp$ 
is given by 
$$\frac{dJ_\perp}{dl}=-J_\perp+ \frac{\lambda^2}{v_F\Lambda(l)}\{\frac{v_F}{2} 
[\frac{(K_c(l)v_c(l))-(K_s(l)v_s(l))}{2}(1+\cos\Lambda(l)R)$$
$$-\frac{(K_c(l)/v_c(l))-(K_s(l)/v_s(l))}{2}(1-\cos\Lambda(l)R)]\}. \eqno(A9) $$  
In Eqs. (A7), (A8) and (A9) $\lambda$ obeys the scaling equation (A4), 
$\Lambda(l)=\Lambda e^{-l}$ and $K_c(l)$, $K_s(l)$, $V_c(l)$ and $V_s(l)$ 
are determined by the sine-Gordon scaling (see \cite{neko}). 

The values of $\hat{J}_\parallel(l_d)$, $\hat{J}_\perp(l_d)$ and 
$\hat{I}_\parallel(l_d)$ are obtained by solving Eqs. (A7)-(A9) with the 
initial conditions $I_\parallel(0)=J_\parallel(0)=J_\perp(0)=0$.  The induced 
coupling constants are given by $\hat{I}_\parallel(l_d)$, $\hat{J}_\parallel(l_d)$ 
and $\hat{J}_\perp(l_d)$.  These couplings are obtained from ${I}_\parallel(l_d)$, 
${J}_\parallel(l_d)$ and ${J}_\perp(l_d)$.  $\hat{I}_\parallel(l_d)= 
{I}_\parallel(l_d)(1/2\pi a)(d/a)$, $\hat{J}_\parallel(l_d)={J}_\parallel(l_d) 
(1/2\pi a)(d/a)$, and $\hat{J}_\perp(l_d)={J}_\perp(l_d)(1/2\pi a)(d/a)$. 
The values $\hat{I}_\parallel(l_d)$, $\hat{J}_\parallel(l_d)$ and 
$\hat{J}_\perp(l_d)$ depend on the sine-Gordon scaling of the short wire. 

Next we consider the two cases: 

(a) $(G-4k_F)d<2\pi$.  As a result the backward term controls the scaling, 
$K_c/v_c\simeq0$, $K_s/v_s\sim(1/2v_F)(1-U/\pi v_F)$.  
$$g_c(l_d)=\frac{U}{2\pi^2}\left(\frac{d}{a}\right)^{(3-2K_c)}\frac{1}{d},~~~K_c<1 .$$ 
Using 
$$\frac{U/a}{\epsilon_0}\simeq\frac{1}{137}\frac{c}{v_F}\frac{d}{a}\frac{1}{k} ,$$ 
we find 
$$\hat{g}_s(l_d)=\frac{\epsilon_0}{k2\pi^2 137}\left(\frac{c}{v_F}\right) 
\left(\frac{d}{a}\right)^{(3-2K_c)}$$ 
with 
$$K_c=\sqrt{\frac{1-U/\pi l v_+}{1+U/\pi l v_+}}\simeq\sqrt{\frac{1-(1/137) 
(c/v_F)(1/\pi k)}{1+(1/137)(c/v_F)(1/\pi k)} } .$$  
Using $g_c(l_d)$ and $K_c$ we obtain the exchange coupling constants. 
$$\hat{J}_\parallel(l_d)=\frac{\lambda^2(0)}{4\pi v_F}\log\left(\frac{d}{a}
\right)\left[1-\frac{U}{2\pi v_F}(1-\langle\cos\Lambda d\rangle)\right] , 
\eqno(A10) $$   
$$\hat{I}_\parallel(l_d)=\frac{\lambda^2(0)}{4\pi v_F}\log\left(\frac{d}{a}
\right)[1+\langle\cos\Lambda d\rangle]    , \eqno(A11) $$ 
and $\hat{J}_\perp(l_d)$ is given by 
$$\hat{J}_\perp(l_d)=\frac{\lambda^2(0)}{2\pi v_F}\log\left(\frac{d}{a}
\right)\left[\left(1-\frac{U}{\pi v_F}\right)(1-\langle\cos\Lambda d\rangle) 
\right] . \eqno(A12) $$ 

In Eqs. (A10)-(A12) $\lambda(0)$ is given by $\lambda(0)=2\sqrt{2}t\sin(k_Fa)$. 
$U$ is the original Hubbard interaction and $\langle\cos\Lambda R\rangle$ 
is given by 
$$\langle\cos\Lambda R\rangle=\frac{1}{l_d}\int_0^{l_d}\cos(\Lambda e^{-l})dl 
=\frac{1}{l_d}\int_{R/d}^{R/a}dx\frac{\cos x}{x}=\frac{1}{\log(d/a)}  
\int_1^{d/a}dx\frac{\cos x}{x}  . \eqno(A13) $$ 
 
(b) For the generic case $(G-4k_F)d>\pi$ we can neglect the Umklapp term. 
We use the relations $K_cv_c=K_sv_s=\tilde{v}_F=2v_F$; 
$$\frac{K_c}{v_c}=\frac{K_c^2}{\tilde{v}_F}=\frac{1}{2v_F}\left(\frac 
{1-U/\pi v_+}{1+U/\pi v_+}\right) $$ 
and 
$$\frac{K_s(l)}{v_s(l)}=\frac{K_s^2}{2v_F}\left[1+\frac{K_s^2}{2v_F}\frac 
{\pi}{2}\int_0^l dl'\hat{g}^2_s(l)\right]^{-1}\simeq\frac{K_s^2}{2v_F} 
-O\left(\frac{K_s^2}{2v_F}\right) . $$ 
As a result we obtain 
$$\frac{K_c}{v_c}+\frac{K_s}{v_s}\simeq\frac{1}{2v_F}\left(1-\frac{2U} 
{\pi v_+}+1+\frac{2U}{\pi v_-}\right)\simeq\frac{1}{v_F} , $$ 
and  
$$\frac{K_c}{v_c}-\frac{K_s}{v_s}\simeq\left(\frac{U}{\pi v_F}\right) 
\frac{1}{v_F} .$$  
Substituting these values into the scaling equations (A7)-(A9) we find: 
$$\hat{J}_\parallel(l_d)=\frac{\lambda^2(0)}{4\pi v_F}\log\left(\frac{d}{a} 
\right)  , \eqno(A14) $$ 
$$\hat{I}_\parallel(l_d)=\frac{\lambda^2(0)}{4\pi v_F}\log\left(\frac{d}{a} 
\right)\langle\cos\Lambda d\rangle  , \eqno(A15) $$ 
$$\hat{J}_\perp(l_d)=\frac{\lambda^2(0)}{2\pi v_F}\left(\frac{U}{4\pi v_F} 
\right)\log\left(\frac{d}{a}\right)[1-\langle\cos\Lambda d\rangle] . 
\eqno(A16) $$

\section{appendix B} 
Here, we consider the boundary term in the leads with the overlapping 
region $\varepsilon\sim a$.
$$H_{BC}=-t_0\sum_{\sigma=\uparrow,\downarrow}[(e^{ik_Fa}L^\dag_\sigma(-a)
-e^{-ik_Fa}L^\dag_\sigma(a))(e^{-ik_F\varepsilon}L_\sigma(-\varepsilon)
-e^{ik_F\varepsilon} L_\sigma(\varepsilon))] $$
$$-[(e^{-ik_Fa}R^\dag_\sigma(a)-e^{ik_Fa} R^\dag_\sigma(-a))(e^{ik_F\varepsilon}
R_\sigma(\varepsilon)-e^{-ik_F\varepsilon} R_\sigma(-\varepsilon))] $$
$$=t_0\sum_{\sigma=\uparrow,\downarrow}\sin(k_Fa)\sin(k_F\varepsilon)
[(L^\dag_{\sigma}(-a)-L^{\dag}_\sigma(a))(L_\sigma(-\varepsilon)-L_\sigma
(\varepsilon)) $$
$$+(R^\dag_\sigma(a)-R^\dag_\sigma(-a))(R_\sigma(\varepsilon)
-R_\sigma(-\varepsilon))]+H.c.  \eqno(B1) $$

The leads Hamiltonian in Eq. (1e) can be written in terms of even and
odd chiral fermions: 
$$\psi_{e,\sigma}(x)=\frac{1}{\sqrt{2}}(R_\sigma(x)+L_\sigma(x)); ~~~~
\psi_{o,\sigma}(x)=\frac{1}{\sqrt{2}}(R_\sigma(x)-L_\sigma(x)).  \eqno(B2)$$
$$H_0^{leads}=\hbar v_F\sum_{\sigma=\uparrow,\downarrow}\int_{-L/2}^{L/2}
dx [\psi^\dag_{e,\sigma}(x)(-i\partial_x)\psi_{e,\sigma}(x)+\psi^\dag_{o,\sigma}
(x)(-i\partial_x) \psi_{o,\sigma}(x)] . \eqno(B3) $$
The boundary term $H_{BC}$ becomes,
$$H_{BC}=\frac{\hbar}{2}(\sin(k_Fa))\sin(k_F\varepsilon)
\sum_{\sigma=\uparrow,\downarrow}\{[(\psi^\dag_{e,\sigma}(-a)
-\psi^\dag_{e,\sigma}(a))
-(\psi^\dag_{o,\sigma}(a)-\psi^\dag_{o,\sigma}(-a))]$$
$$[(\psi_{e,\sigma}(-\varepsilon)
-\psi_{e,\sigma}(\varepsilon))-(\psi_{o,\sigma}(-\varepsilon)-\psi_{o,\sigma}
(\varepsilon))]+[(\psi^\dag_{e,\sigma}(a)-\psi^\dag_{e,\sigma}(-a))+
(\psi^\dag_{o,\sigma}(a)-\psi^\dag_{o,\sigma}(-a))]$$
$$[(\psi_{e,\sigma}(\varepsilon)
-\psi_{e,\sigma}(-\varepsilon))+(\psi_{o,\sigma}(\varepsilon)-\psi_{o,\sigma}
(-\varepsilon))]\} + H.c.  \eqno(B4) $$

The representation (1i) is used in the Kondo regime where the
{\it symmetric state $\psi_{o,\sigma}(x)$ is screened out} and $H_{BC}$
is replaced by the {\it antisymmetric state $\psi_{e,\sigma}(x)$}.
In the Kondo limit $H_{BC}$ is replaced by a ``mass" term [the
designation symmetric or antisymmetric is given according to the
original fermions $c_{R,\sigma}(x)$ and $c_{L,\sigma}(x)$],
$$H_{BC}\simeq-2t_0(2\sin(k_Fa))^2\sum_{\sigma=\uparrow,\downarrow}
\psi^\dag_{e,\sigma}(0)\psi_{e,\sigma}(0).  \eqno(B5)$$

\begin{figure}[t] 
\caption{Conductance ($G$) as a function of gate voltage ($\hat{U}_G$) 
for fixed weak transmission ($\Gamma^2=0.1$) for three different 
temperatures: $\beta=1$ (dotted line), $\beta=10$ (dashed line) and 
$\beta=50$ (solid line).  Note that decreasing the temperature to 
$\beta=10$ leads to the conductance reaching the value of 1.0 and in 
addition there is a shoulder around 0.7. } 
\end{figure}

\begin{figure}
\caption{Conductance as a function of gate voltage for large 
transmission ($\Gamma^2=5.0$), i.e., for a short wire in the Kondo regime 
at very low temperature ($\beta=100$).  The conductance is about 
$1.0\times2e^2/\hbar$, as expected from the Kondo solution. } 
\end{figure}

\end{document}